\documentclass[preprint,pra,showpacs,showkeys,floatfix]{revtex4}

\usepackage{graphicx}
\usepackage{dcolumn}
\usepackage{array}
\usepackage{amsmath}
\usepackage{amsfonts}

\begin{document}

\title{Supersymmetry and eigensurface topology \\of the spherical quantum pendulum}

\author{Burkhard Schmidt}
\email{burkhard.schmidt@fu-berlin.de}
\affiliation{
Institute for Mathematics, Freie Universit\"{a}t Berlin \\ Arnimallee 6, D-14195 Berlin, Germany}

\author{Bretislav Friedrich}
\email{bretislav.friedrich@fhi-berlin.mpg.de}
\affiliation{
Fritz-Haber-Institut der Max-Planck-Gesellschaft \\ Faradayweg 4-6, D-14195 Berlin, Germany}

\date{\today}

\begin{abstract}

 We undertook a mutually complementary analytic and computational study of the full-fledged spherical (3D) quantum rotor subject to combined orienting and aligning interactions, corresponding to a linear polar and polarizable molecule interacting with an electric field. The orienting and aligning interactions are characterized, respectively, by dimensionless parameters $\eta$ and $\zeta$. By making use of supersymmetric quantum mechanics (SUSY QM), we found two sets of conditions (Cases A and B) under which the problem of a spherical quantum pendulum becomes analytically solvable. These conditions coincide with the loci $\zeta=\frac{\eta^2}{4k^2}$ of the intersections of the eigenenergy surfaces spanned by the $\eta$ and $\zeta$ parameters. In Case A, the topological index $k=\mp m+1$ with $m$ the projection quantum number, whereas in Case B, $k=1$, independent of $m$. These findings have repercussions for rotational spectra and dynamics of molecules subject to combined permanent and induced dipole interactions.

\end{abstract}


\maketitle

\section{Introduction}

The problem of the spherical quantum pendulum, i.e.,  that of a 3D rigid rotor under a cosine potential and/or its variants, belongs to prototypical problems in quantum mechanics. It
lurks behind numerous  applications in molecular physics, such as orientation/alignment of molecules \cite{SlenFriHerPRL1994,FriSlenHer1994,JCP1999FriHer, FriHerJPCA99, Ortigoso1999,Seideman:99a,Seideman1999,Larsen:00a,CaiFri2001,CCCC2001,Averbukh:01a, Leibscher:03a,PRL2003Sakai, JMO2003-Fri,Leibscher:04a,Buck-Farnik,HaerteltFriedrichJCP08,PotFarBuckFri2008,Leibscher:09a,PRL2009Stap, Owschimikow2009,Ohshima2010,Averbukh2010,Owschimikow2010,Owschimikow2011}, deflection and focusing of molecular translation \cite{ZhaoFriChung2003,JCP2009Kuepp}, molecular trapping \cite{ChemRev2012-Meijer}, attaining time-resolved photoelectron angular distributions \cite{Holmegaard2010,Bisgaard2009,Hansen2011}, diffraction-from-within \cite{Landers2001}, separation of photodissociation products \cite {IsraelJ2003-Manz,JCP2004-Manz, Lorenz:11a}, deracemization \cite {JCP2009Stapelfeldt}, high-order harmonic generation and orbital imaging \cite{Itatani:04a,CorkumHHG, BandraukHHG, IvanovHHG, Smirnova2009,Woerner,Mohn2012}, quantum simulation \cite{Bar:12,Man:13} or quantum computing \cite{DeMille2002,de6,de7,de8,QiKaisFrieHer2011a,QiKaisFrieHer2011b}, see also Ref. \cite{LemKreDoyKais:MP2013}. Herein, we  revisit the fundamentals of the spherical pendulum's quantum mechanics with an eye on the physics of a linear polar and polarizable molecule interacting with an electric field (combined permanent and induced electric dipole interaction). 

Compared with other prototypical problems in quantum mechanics, that of the spherical quantum pendulum is of a recent date \cite{Schlier1955,Meyenn1970,FriHer1991ZPhys,FriHerPRL1995}. A general solution of the full-fledged (3D) pendular eigenproblem relies on numerical diagonalization of a truncated infinite-dimensional Hamiltonian matrix, whose elements can be conveniently expressed in the free-rotor basis set. The matrix elements are a function of dimensionless parameters $\eta$ and $\zeta$ that characterize, respectively, the strengths of the pendulum's orienting ($\propto \eta \cos\theta$) and aligning ($\propto \zeta \cos^2\theta$) interactions (where $\theta$ is a polar angle). These interactions correspond to those of a polar (permanent electric dipole moment) and polarizable (induced electric dipole moment) molecule subject to either an electrostatic field or a combination of an electric and an far-off-resonant optical field or a single-cycle non-resonant optical field \cite{JCP1999FriHer,HaerteltFriedrichJCP08,Leibscher:09a,Owschimikow2009}. For arbitrary values of the $\eta$ and $\zeta$ parameters, there are no analytic, closed-form solutions to the pendular eigenproblem. 

In our previous work \cite{LemMusKaisFriPRA2011,LemMusKaisFriNJP2011}, we showed that the pendular eigenproblem can in fact be solved analytically, but only for a particular ratio of $\eta$ to $\zeta$. We found this ratio -- which represents a condition for analytic solvability -- as well as the analytic solution itself by invoking the apparatus of supersymmetric quantum mechanics (SUSY QM) \cite{s4,s36,s24,s27,s26,s17}. 

In follow-up work \cite{SchmiFri2014a}, we analyzed the intersections of the eigenenergy surfaces (eigensurfaces) spanned by the  $\eta$ and $\zeta$ parameters and found that their loci can be rendered analytically and that the eigenstates as well as the number of their intersections can be characterized by a single integer index (termed topological index), $k\equiv\frac{\eta}{2\sqrt{\zeta}}$, distinctive for a particular ratio of the interaction parameters $\eta$ and $\zeta$. Moreover, we found that the topological index is connected with the condition for exact solvability as established by SUSY: the exact solutions that we found only obtained for the value of the topological index $k=-m+1$, with  $m$ the angular momentum projection quantum number. In the next step \cite{SchmiFri2014b}, we established that the reduced-dimensionality analog, the planar pendulum problem, has at least three classes of analytic solutions and that each  is characterized by a particular value of the pertinent topological index. 

In this paper, we return to the spherical rotor/pendulum problem, but make use of a different {\it Ansatz} for the superpotential. This {\it Ansatz} yields not only the class of solutions to the pendular eigenproblem obtained previously but also a new class of analytic solutions. Like for the class of solutions found before, the new class can again be characterized by a topological index. Its value is, however, always $k=1$, independent of $m$.

This paper is organized as follows: 
In Sec. \ref{sec:Hamiltonian} we introduce the Hamiltonian of the spherical pendulum along with the dimensionless parameters $\eta$ and $\zeta$. In Sec. \ref{sec:Eigenproperties} we present, in turn, the cases of a purely orienting interaction, purely aligning interaction and of a combined orienting and aligning interaction and identify a condition for the (avoided) intersections of the eigensurfaces. 
Moreover, we show that the topology of the intersections can be characterized by a single topological index. 
In Sec. \ref {sec:susy}, we present two sets of conditions that lead to analytic solutions of the quantum pendulum eigenproblem and find that these sets pertain to a particular topological index.
In Sec. \ref{sec:examples} we discuss examples of insights that can be gained from SUSY for choice values of the quantum number $m$ and the topological index $k$. 
In Sec. \ref{sec:outlook}, we provide a summary of the present work.

\section{The Hamiltonian}
\label{sec:Hamiltonian}

We consider a spherical quantum pendulum (particle on a sphere) problem whose Hamiltonian is given by 
\begin{equation}
\mathcal{H} = B\mathbf{J}^2 +BU(\theta)
\label{eq:H}
\end{equation}
where the  kinetic energy term is given by the square of the angular momentum operator
\begin{equation}
\mathbf{J}^2 = - \frac{1}{\sin\theta} \frac{\partial}{\partial\theta}\left(\sin\theta\frac{\partial}{\partial\theta}\right)
- \frac{1}{\sin^2\theta}\frac{\partial^2}{\partial\phi^2}
\label{eq:J2}
\end{equation}
and where the potential
\begin{equation}
U(\theta) = - \eta \cos \theta - \zeta \cos^2 \theta
\label{eq:pot3D}
\end{equation}
is restricted to the lowest two Fourier terms in the polar angle, $0 \le \theta \le \pi$. In what follows, we will express all energies in terms of the rotational constant, $B$, which amounts to setting $B=1$ in Eq. (\ref{eq:H}) \cite{SchmiFri2014a}. Since the $\cos\theta$ and $\cos^2\theta$ terms generate, respectively, oriented (single arrow-like) and aligned (double-arrow-like) states, we term the two interactions {\it orienting} and {\it aligning}. Their strengths are characterized, respectively, by the dimensionless parameters $\eta>0$ and $\zeta>0$. The relationships of the dimensionless parameters  to the properties of molecules and fields are given in Subsections \ref{sec:eta} and \ref{sec:zeta}. For  $\eta=\zeta=0$, Hamiltonian (\ref{eq:H}) becomes that of a free rotor.

Since Hamiltonian (\ref{eq:H}) is independent of the azimuthal angle, $\phi$, the corresponding Schr\"{o}dinger equation, $\mathcal{H}\chi(\theta,\phi)=E\chi(\theta,\phi)$, can be conveniently solved by making use of the  Ansatz
\begin{equation}
  \chi(\theta,\phi)=(\sin\theta)^{-\frac{1}{2}}\psi(\theta)e^{im\phi}
	\label{chi}
\end{equation}
where $m$ is a good (integer) quantum number. The wave function $\psi(\theta)$ satisfies the 
 one-dimensional Schr\"{o}dinger equation in the polar angle only, $H \psi(\theta)=E\psi(\theta)$, with  Hamiltonian
\begin{equation}
H  = -\frac{d^2}{d\theta^2} + V (\theta)
\label{eq:hamilton}
\end{equation}
where 
\begin{eqnarray}
V (\theta) &=& \left(m^2-\tfrac{1}{4}\right) \csc^2 \theta  + V(\theta) -\tfrac{1}{4} \nonumber \\
&=& \left(m^2-\tfrac{1}{4}\right) \csc^2 \theta - \eta \cos \theta - \zeta \cos^2 \theta -\tfrac{1}{4} 
\label{eq:pot}
\end{eqnarray}
is an effective one-dimensional potential. We note that the wavefunction $\psi(\theta)$ has a unit Jacobian factor. The spherical pendulum's potential in the form of Eq. (\ref{eq:pot}) will be considered throughout the remainder of this paper.

\section{Eigenproperties}
\label{sec:Eigenproperties}

\subsection{Pure orienting interaction:  $\eta > 0$ \& $\zeta=0$ }
\label{sec:eta}

When a rigid rotor interacts with an external field solely via the orienting interaction,  the  eigenproblem for the corresponding operator $\mathcal{H}_{\eta}\equiv \mathbf{J}^2-\eta \cos\theta$ can be conveniently solved by diagonalizing the Hamiltonian matrix $\{\langle j',m'|\mathcal{H}_{\eta}|j,m\rangle\}$ in the free-rotor basis set $|j,m\rangle$, with $j\ge 0$ and $-j\le m \le j$. Its matrix elements are given by
\begin{eqnarray}
 \langle j',m'|\mathcal{H}_{\eta}|j,m\rangle= & j(j+1)\delta_{j',j}\delta_{m',m} -\eta \left[\frac{(j+m)(j-m)}{(2j+1)(2j-1)}\right]^{\frac{1}{2}} \delta_{j',j-1}\delta_{m',m} \nonumber \\
 & -\eta \left[\frac{(j+m+1)(j-m+1)}{(2j+3)(2j+1)}\right]^{\frac{1}{2}} \delta_{j',j+1}\delta_{m',m}
 \label{eq:matrixelemeta}
 \end{eqnarray}
 
As the $\cos\theta$ operator mixes basis states of opposite parity, the resulting hybrid wave functions, $|J,m;\eta\rangle$, are of {\it indefinite} parity. The hybrid states are labeled by the values of $J$ the states would have in the free-rotor limit, $|J,m;\eta\rightarrow0\rangle\rightarrow |j,m\rangle$, and by the good quantum number $|m|$. All states bound by the attractive single-well $-\eta\cos\theta$ potential occur as singlets.

\begin{figure}
  \centering
  \includegraphics[width=12cm]{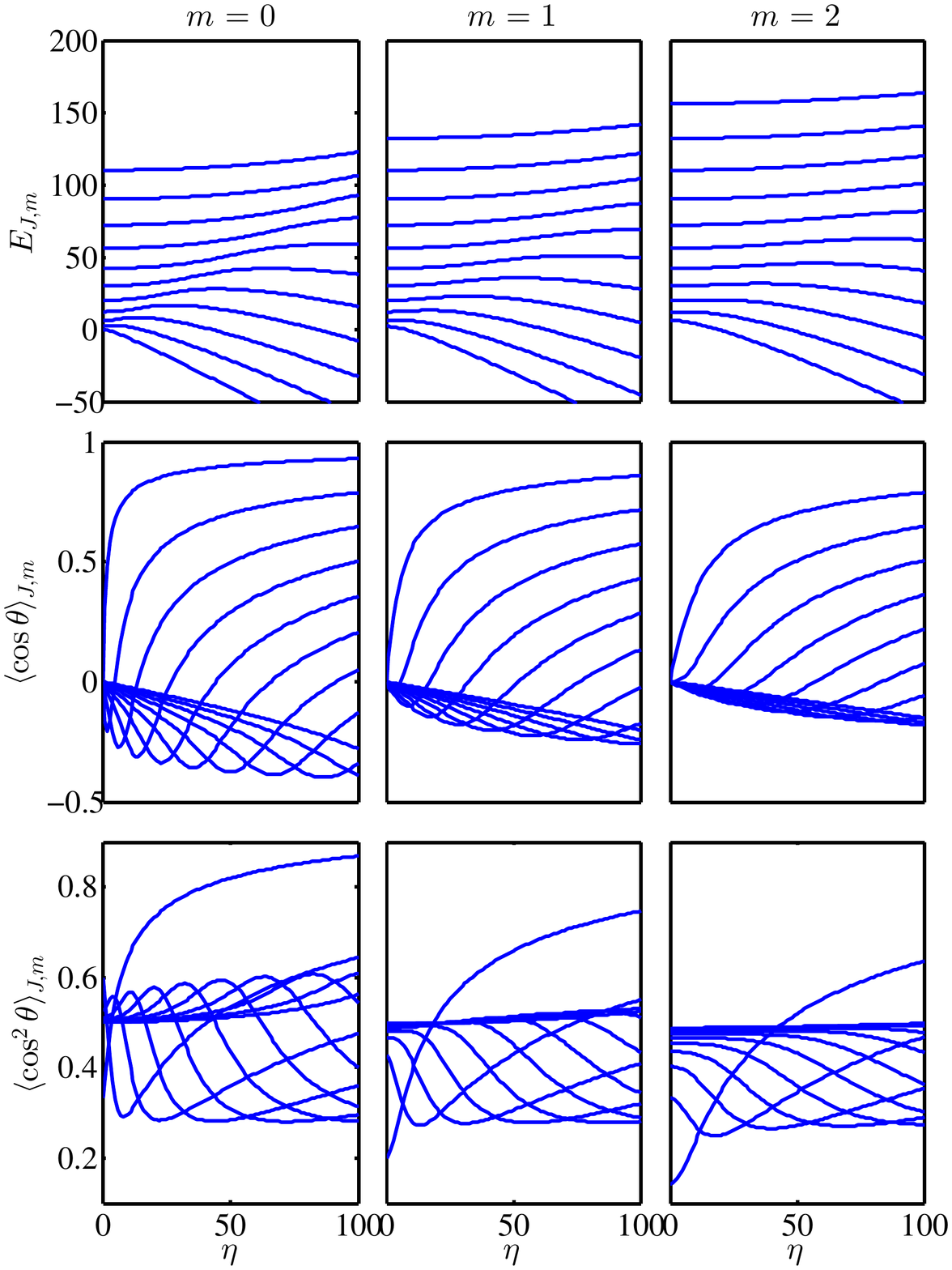}
  \caption{(Color online) Eigenenergies (expressed in units of the rotational constant, B, and ordered from bottom to top according to the increasing value of $\tilde{J}$) and the corresponding orientation and alignment cosines for a spherical rotor subject to purely orienting interaction ($\zeta=0$).}
  \label{fig:eta3}
\end{figure}

 Figure \ref{fig:eta3} shows the eigenenergies, $E_{J,|m|}$ and orientation, $\langle\cos\theta\rangle_{J,|m|}$, and alignment, $\langle\cos^2\theta\rangle_{J,|m|}$, cosines for the lowest states. One can see that only for the ``stretched'' states, with $J=|m|$, that are the lowest energy states, are the dependencies of the eigenproperties on $\eta$ monotonous (there are no states beneath to ``repel'' them). The rest is subject to the ``Stern effect'' \cite{FriHer1991ZPhys,Friedrich:91b}, where the states become first anti-oriented/anti-aligned at low $\eta$, before conforming to the direction of the orienting field at large $\eta$ \cite{SlenFriHer1994,EPJD2006_NetPolarization}. 
 
Only in the free-rotor, $\eta\rightarrow0$, and the harmonic-librator, $\eta\rightarrow\infty$, limit are the eigenenergies (and other eigenproperties)  analytic, given, respectively, by 
\begin{equation}
E(\eta=0)=j(j+1) 
\end{equation}
and
\begin{equation}
E(\eta\rightarrow \infty)=-\eta + n\sqrt{2\eta}
\end{equation}
with $j=0,1,2,...$ and $n=2J-|m|$, cf. Refs. \cite{Meyenn1970, RostFriHerPRL1992,IRPC1996FriHer}.

We note that for a polar $^{1}\Sigma $ rigid-rotor molecule with a
body-fixed dipole moment $\mu $ and a rotational constant $B$ subject to
an electric field, $\boldsymbol{\varepsilon }_1$, the orienting  parameter is  given by 
\begin{equation}
\eta \equiv \frac{\mu \varepsilon_1}{B}
\label{eq:defeta}
\end{equation}
The electric vector $\boldsymbol{\varepsilon}_{1}$ can be due to an electrostatic field or a temporally shaped optical field.

\subsection{Pure aligning interaction:  $\eta = 0$ \& $\zeta>0$ }
\label{sec:zeta}

When a rigid rotor interacts with an external field solely via the aligning interaction,  the  eigenproblem for the corresponding operator $ \mathcal{H}_{\zeta}\equiv \mathbf{J}^2-\zeta \cos^2\theta$ can be conveniently solved by diagonalizing the Hamiltonian matrix $\{\langle j',m'| \mathcal{H}_{\zeta}|j,m\rangle\}$ in the free-rotor basis set $|j,m\rangle$. Its matrix elements are given by
\begin{eqnarray}
 \langle j',m'| \mathcal{H}_{\zeta}|j,m\rangle=& j(j+1)\delta_{j',j}\delta_{m',m}-\zeta \left[\frac{1}{3}+\frac{2(2j+1)[j(j+1)-3m^2]}{3(2j+3)(2j+1)(2j-1)}\right] \delta_{j',j}\delta_{m',m} \nonumber \\
 & -\zeta\left[\frac{[(j+m)(j+m-1)(j-m)(j-m-1)]}{(2j-1)^2(2j+1)(2j-3)}\right] ^{\frac{1}{2}}\delta_{j',j-2}\delta_{m',m}\nonumber \\
 & -\zeta\left[\frac{[(j+m+2)(j+m+1)(j-m+2)(j-m+1)]}{(2j+3)^2(2j+5)(2j+1)}\right] ^{\frac{1}{2}}\delta_{j',j+2}\delta_{m',m}
 \label{eq:matrixelemzeta}
 \end{eqnarray}
Since the $\cos^2\theta$ operator mixes basis states of same parity, the resulting hybrid wave functions, $|J,m;\zeta\rangle$, are of {\it definite} parity.
Furthermore, all states bound by the attractive $-\zeta\cos^2\theta$ potential occur as doublets split by tunneling through the potential's equatorial barrier, whereby the members of a given tunneling doublet have same $|m|$ but opposite parity.

The Schr\"{o}dinger equation for Hamiltonian $ \mathcal{H}_{\zeta}$ is isomorphic with the oblate spheroidal wave equation, cf. Refs. \cite{Abramovitz:72a,FriHerPRL1995,FriHerJPC95}. Only in the free-rotor, $\eta\rightarrow0$, and the harmonic-librator, $\eta\rightarrow\infty$, limit are the corresponding eigenenergies (and other eigenproperties)  analytic.  The harmonic librator eigenenergies are given by  
\begin{eqnarray}
& E_{\text{even}}(J,m;\zeta \rightarrow \infty)= -\zeta + 2\sqrt{\zeta}+2J\sqrt{\zeta}+\frac{m^2}{2}-\frac{J^2}{2}-J-\frac{1}{2} \hspace{.2cm}\text{for} \hspace{.2cm}(J-m) \hspace{.2cm}\text{even} \nonumber \\
& E_{\text{odd}}(J',m';\zeta \rightarrow \infty)=-\zeta + 2J\sqrt{\zeta}+\frac{m'^2}{2}-\frac{J'^2}{2}-\frac{1}{2} \hspace{.2cm} \text{for}\hspace{.2cm} (J'-m') \hspace{.2cm}\text{odd}
\end{eqnarray}
cf. also Refs.  \cite{FriHerZPhys96,IRPC1996FriHer,EPJD2006_NetPolarization}. 
These are equal for the members of a given tunneling doublet, $E_{\text{even}}(J,m;\zeta \rightarrow \infty) =E_{\text{odd}}(J+1,m;\zeta \rightarrow \infty)$ or $E_{\text{odd}}(J,m;\zeta \rightarrow \infty) =E_{\text{even}}(J+1,m;\zeta \rightarrow \infty)$. Successive tunneling doublets are separated by an energy gap of  $2\zeta^{\frac{1}{2}}$.

\begin{figure}
  \centering
  \includegraphics[width=12cm]{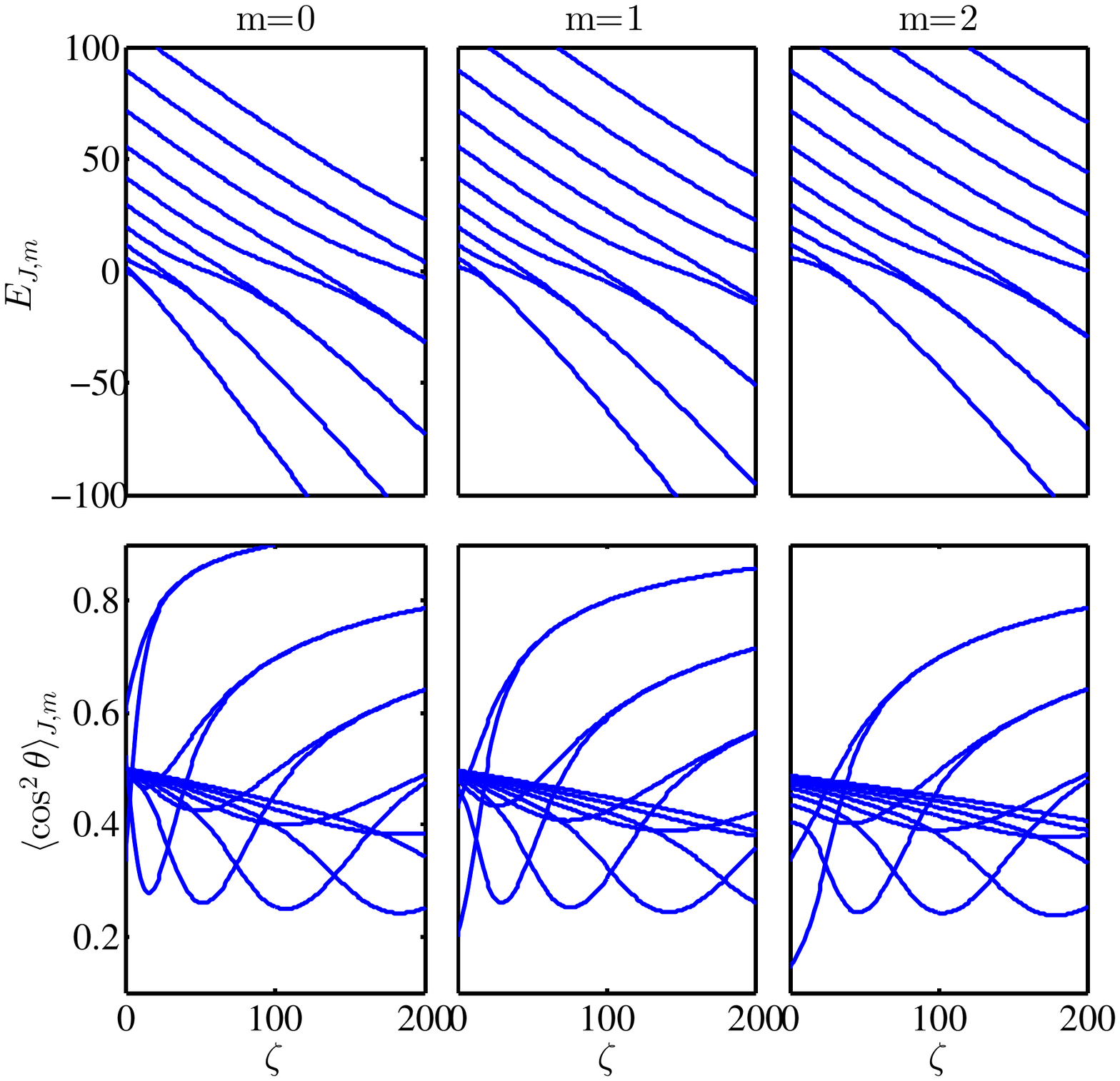}
  \caption{(Color online) Eigenenergies (expressed in units of the rotational constant, B, and ordered from bottom to top according to the increasing value of $\tilde{J}$) and the corresponding alignment cosine for a  spherical rotor subject to purely aligning interaction ($\eta=0$). One can see the formation of the tunneling doublets whose eigenenergies become quasi-degenerate in the limit of large $\zeta$ and the corresponding alignment cosines coincide as a result. }
  \label{fig:zeta3}
\end{figure}

Figure \ref{fig:zeta3} shows the eigenenergies, $E_{J,|m|}$ and alignment cosines, $\langle\cos^2\theta\rangle_{J,|m|}$,  for the lowest states labeled by the values of $J$ the states would have in the free-rotor limit, $|J,m;\eta\rightarrow0\rangle\rightarrow |j,m\rangle$, and by the good quantum number $|m|$. One can see that only for the lowest tunneling doublets, with $J=|m|$ and $J=|m|+1$, are the dependencies of the alignment cosines on $\zeta$ monotonous (there are no states beneath to ``repel'' them). The rest is again subject to the ``Stern effect,'' where the states become first anti-aligned at low $\zeta$, before conforming to the direction of the aligning field at large $\zeta$ \cite{FriHerJPC95,EPJD2006_NetPolarization}. 

We note that for a polarizable $^{1}\Sigma $ rigid-rotor molecule with body-fixed static-polarizability components 
$\alpha _{||}$ and $\alpha _{\bot }$, and a rotational constant $B$ subject
to an electric field, $\boldsymbol{\varepsilon }_2$, the aligning parameter is  given by 
\begin{equation}
\zeta \equiv \zeta
_{||}-\zeta _{\bot }\text{\hspace{1cm}}\zeta _{||,\bot }\equiv \frac{%
\alpha _{||,\bot }\varepsilon_2^{2}}{2B}.
\label{eq:defzeta}
\end{equation}
The electric vector $\boldsymbol{\varepsilon}_{2}$ can be due to an electrostatic field or to a non-resonant optical field of intensity $I$ such that 
\begin{equation}
\varepsilon_2 =\left( \frac{2 I}{c\epsilon_0}\right) ^{1/2}
\end{equation}
with $c$ the speed of light in vacuum and $\epsilon_0$ the vacuum permittivity. 

\subsection{Combined orienting and aligning interactions:  $\eta > 0$ \& $\zeta>0$}
\label{sec:combined}

When a rigid rotor interacts with an external field via a combined orienting and aligning interaction,  the  eigenproblem for the corresponding operator, given by Eq. (\ref{eq:H}),  can be conveniently solved by diagonalizing the Hamiltonian matrix $\{\langle j',m'| \mathcal{H}|j,m\rangle\}$ in the free-rotor basis set $|j,m\rangle$. Its matrix elements are a sum of the matrix elements  $\langle j',m'| \mathcal{H}_{\eta}|j,m\rangle$ and  $\langle j',m'| \mathcal{H}_{\zeta}|j,m\rangle$ given by Eqs. (\ref{eq:matrixelemeta}) and (\ref{eq:matrixelemzeta}), less once the free-rotor $j(j+1)$ term. The resulting hybrid states, $|J,m;\eta,\zeta\rangle$, correlate with the free-rotor states in the field-free limit, $|J,m;\eta\rightarrow0,\zeta\rightarrow0\rangle\rightarrow |J,m\rangle$, and with the harmonic librator states, $|J,m;\eta\rightarrow\infty,\zeta\rightarrow\infty\rangle$, in the strong-field limit, where their equidistant eigenenergies become $-\eta-\zeta +\sqrt{2\eta+4\zeta}(2n\pm m+\frac{1}{2})$, with $n=0,1,2,...$

As shown in our previous work \cite{JCP1999FriHer, FriHerJPCA99, JMO2003-Fri}, the opposite-parity members of the tunneling doublets are coupled by the $\cos\theta$ operator and the corresponding energy levels ``repel'' each other proportionately to the parameter $\eta$. This leads to intersections of the eigensurfaces, $E(\eta,\zeta)$, since the upper member of a lower tunneling doublet is bound to meet the lower member of an upper tunneling doublet. However, all the intersections arising for the spherical pendulum eigenproblem are avoided, since the intersecting opposite-parity states get coupled by the parity-mixing $\cos\theta$ operator.

Following a similar argument as that in Ref. \cite{SchmiFri2014a}, we find that the eigensurfaces $E(\eta,\zeta)$ exhibit  crossings at loci given by 
\begin{equation}
 \frac{\eta}{2\sqrt{\zeta}}=k
\label{eq:k}
\end{equation}
which coincides with Eq. (16) of  Ref. \cite{SchmiFri2014a}. Hence the intersection loci are independent not only of  $J$ but also of $|m|$ and only depend on $k$, the topological index. 

\begin{figure}
  \centering
  \includegraphics[width=12cm]{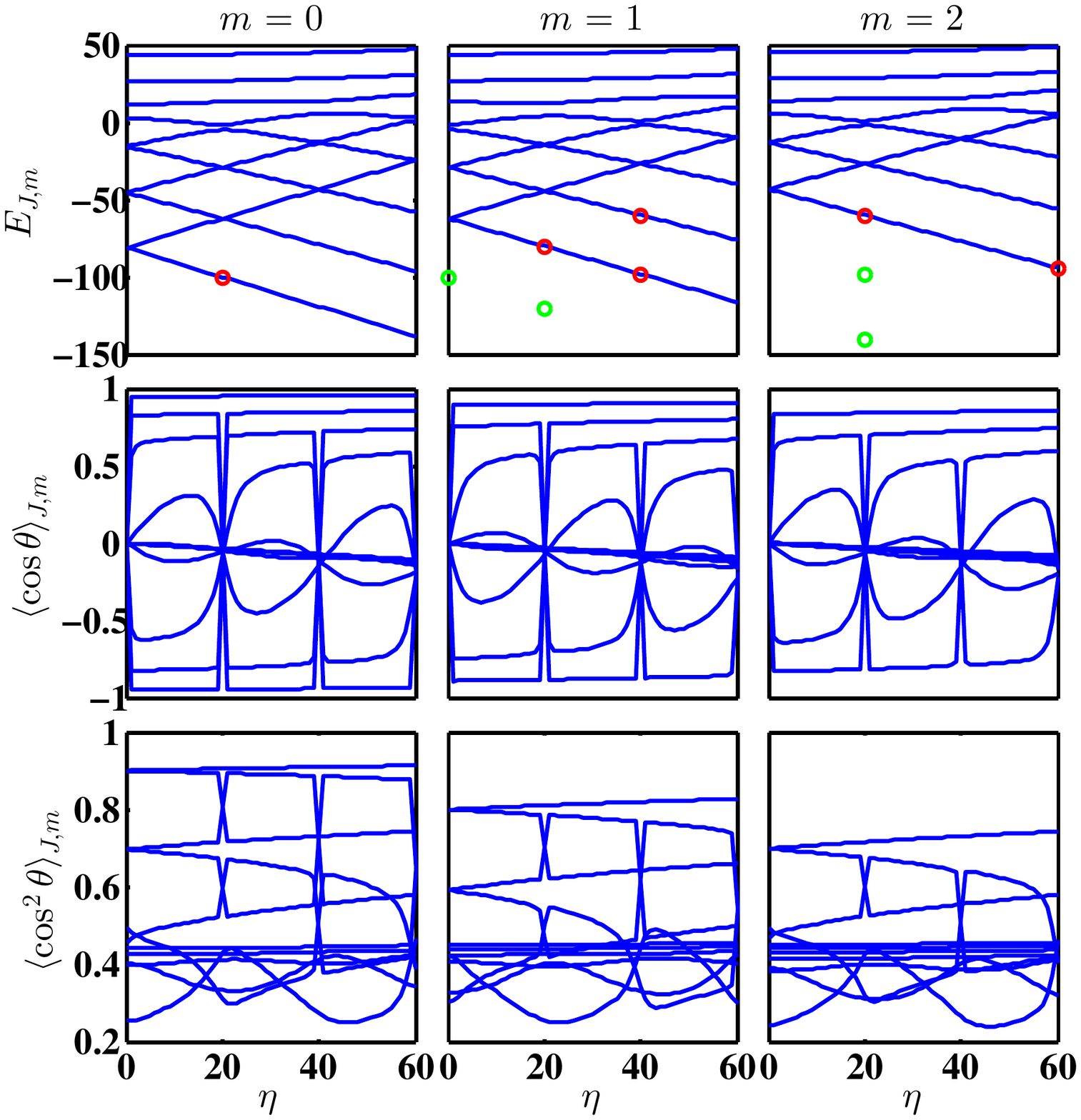}
  \caption{(Color online) Eigenenergies (expressed in units of the rotational constant, $B$) of a spherical rotor subject to the combined orienting and aligning interactions as a function of $\eta$ for a fixed $\zeta=100$. The labeled $\eta$-axis tick marks indicate the positions of the avoided crossings of the energy levels. The symbols (red: normalizable, green: non-normalizable) indicate analytic solutions found via SUSY QM, see Sec. \ref{sec:susy}.}
  \label{fig:cut3}
\end{figure}

The upper panels of Fig. \ref{fig:cut3} show the eigenenergies of a spherical rotor subject to the combined orienting and aligning interactions as a function of $\eta$ for a fixed $\zeta=100$. These exhibit avoided crossings at $\eta$ corresponding to integer values of the topological index $k$ (i.e., at integer multiples of 20 for the example shown), as implied by Eq. (\ref{eq:k}). The lower panels show the dependence on $\eta$ for $\zeta=100$ of the corresponding directional characteristics of the ten lowest states -- the expectation values of the orientation, $\langle\cos\theta\rangle$, and alignment, $\langle\cos^2\theta\rangle$, cosines. The orientation cosines exhibit $J+|m|$ sign changes for each $|J,|m|\rangle$ state, which occur abruptly  at the avoided intersections. Likewise, the alignment cosines exhibit abrupt changes at the avoided intersections. Particularly noteworthy is the high degree of orientation/alignment at even tiny values of $\eta$ at sizable $\zeta$ that arises thanks to the facile coupling of the quasi-degenerate tunneling doublet members (``pseudo first-order Stark effect'' \cite{JCP1999FriHer,JMO2003-Fri}).
 
\begin{figure}
  \centering
  \includegraphics[width=12cm]{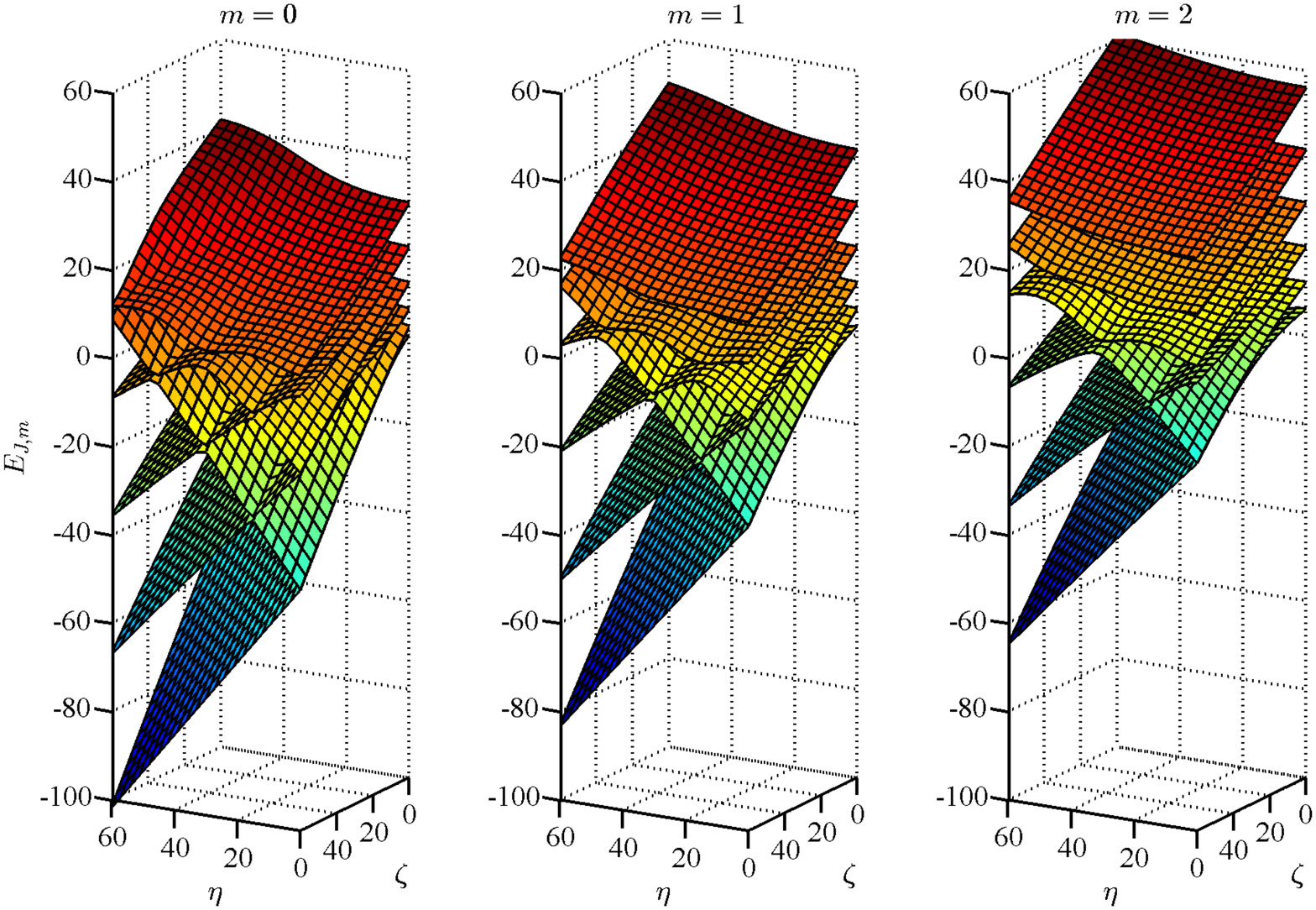}
  \caption{(Color online) Eigensurfaces of a spherical rotor subject to the combined orienting and aligning interactions as a function of $\eta$ and $\zeta$. All intersections are avoided but become quasi-degenerate for large values of $\eta$ and $\zeta$.}
  \label{fig:ene3}
\end{figure}

 Figure \ref{fig:ene3} shows the eigenenergy surfaces spanned by the parameters $\eta$ and $\zeta$ pertaining to the lowest six eigenstates of a spherical rotor subject to the combined interactions. 
As one can see, at $\zeta=0$ or $\eta=0$, the energy surfaces correspond to the purely orienting or purely aligning interactions described above and shown in Figs. \ref{fig:eta3} and \ref{fig:zeta3}. 
As Eq. (\ref{eq:k}) is state-independent, the number of intersections an energy surface partakes in is equal to the label $J$ of the corresponding eigenstate, cf. also upper panel of Fig \ref{fig:cut3}: 
the lowest energy surface, with $J=|m|$, is thus not involved in any intersection; 
the first excited state surface, with $J=|m|+1$, is involved in a first-order ($k=1$) intersection (between nearest doublets); 
the second excited state surface, with $J=|m|+2$, is involved both in a first-order ($k=1$)  intersection (between nearest doublets) and in the second-order ($k=2$) avoided intersection (between second nearest doublets), etc. Consequently, at the loci of the $k$-th order intersections given by Eq.~(\ref{eq:k}), we find an energy level pattern with $k$ single states at the bottom, followed by all other states which are doubly degenerate. In contrast, there are no degeneracies arising anywhere in between these intersection loci.

\begin{figure}
  \centering
  \includegraphics[width=12cm]{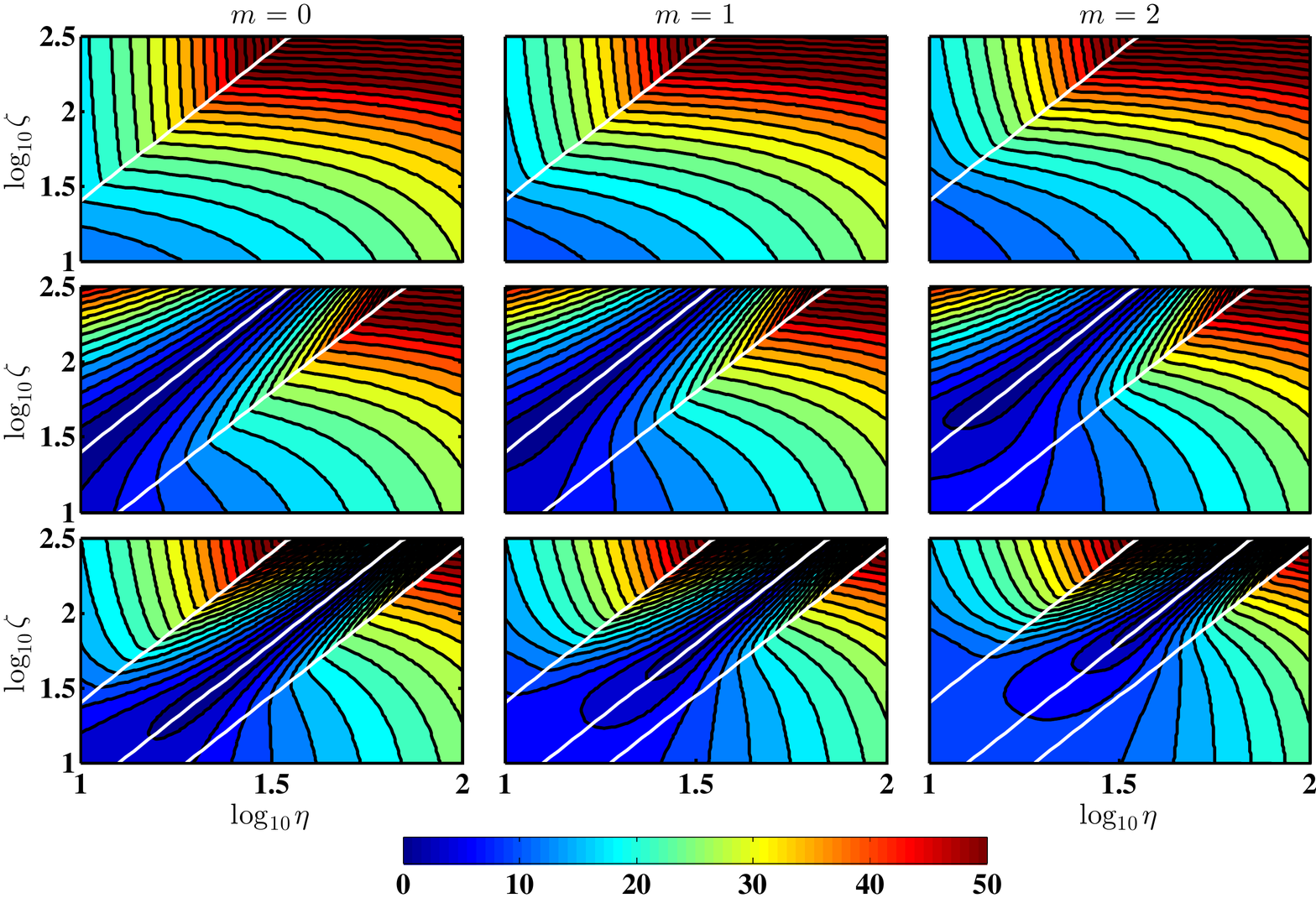}
  \caption{(Color online) Gaps between eigensurfaces of a spherical rotor subject to the combined orienting and aligning interactions as a function of $\eta$ and $\zeta$, both plotted on a logarithmic  (base 10) scale. The gaps help to visualize the intersections of the eigensurfaces seen  in Fig. \ref{fig:ene3}. The white lines show the loci of the eigensurface intersections, cf. Eq. (\ref{eq:k}). The gaps shown pertain to energy differences of states with $J=m$ and $J=m+1$ (top), $J=m+1$ and $J=m+2$ (middle) and $J=m+2$ and $J=m+3$ (bottom).}
  \label{fig:gap3}
\end{figure}

The intersections of the eigenenergy surfaces are visualized in Fig. \ref{fig:gap3} which shows the energy differences (gaps) between adjacent eigenenergy surfaces. The valleys and the ridges occur along straight lines with a slope of 2 in the double-logarithmic scale of the figure, thus conforming to a quadratic dependence of $\zeta$ on $\eta$. The valleys coincide  accurately with the white lines drawn at 
\begin{equation}
\zeta=\frac{1}{4k^2}\eta^2
\label{eq:zeta-eta}
\end{equation}
in agreement with Eq. (\ref{eq:k}).

We note that for a polar and polarizable $^{1}\Sigma $ rigid-rotor molecule with a
body-fixed dipole moment $\mu $, body-fixed static-polarizability components 
$\alpha _{||}$ and $\alpha _{\bot }$, and a rotational constant $B$ subject
to collinear electric fields $\boldsymbol{\varepsilon }_1$ and $\boldsymbol{\varepsilon }_2$, the orienting and aligning parameters are  given by 
Eqs. (\ref{eq:defeta}) and (\ref{eq:defzeta}).
In case the electric vector $\boldsymbol{\varepsilon}_{1}$ is due to an electrostatic field and the vector $\boldsymbol{\varepsilon}_{2}$ to a non-resonant optical field, the fields $\boldsymbol{\varepsilon}_{1}$ and $\boldsymbol{\varepsilon}_{2}$ would act on the permanent and induced dipoles separately, without adding up to a single effective field. However, the induced and permanent dipole interactions can also arise due to the same field $\boldsymbol{\varepsilon}_{1}=\boldsymbol{\varepsilon}_{2}=\boldsymbol{\varepsilon}$, in which case the ratio of the permanent dipole interaction squared to the polarizability interaction is field-independent and fixed for a given molecule \cite{SchmiFri2014a}.

\section{Supersymmetry (SUSY)}
\label{sec:susy}

\subsection{SUSY apparatus}
\label{sec:susyapp}

Supersymmetric quantum mechanics \cite{s17,s26} is based on the 
concept of superpartner Hamiltonians $H_1$ and $H_2$ with corresponding Schr{\"o}dinger equations
\begin{eqnarray}
H_1 \psi_n^{(1)} &=& (A^\dagger A+\epsilon) \psi_n^{(1)} = E_n^{(1)} \psi_n^{(1)} \nonumber \\
H_2 \psi_n^{(2)} &=& (A A^\dagger+\epsilon) \psi_n^{(2)} = E_n^{(2)} \psi_n^{(2)}
\label{eq:TISE12}
\end{eqnarray}
where the symmetry of the construction ensures that  
\begin{eqnarray}
H_1 (A^\dagger \psi_n^{(2)}) &= (A^\dagger A A^\dagger+\epsilon A^\dagger)\psi_n^{(2)} =& E_n^{(2)} (A^\dagger \psi_n^{(2)} ) \nonumber \\
H_2 (A         \psi_n^{(1)}) &= (AA^\dagger A+\epsilon A)\psi_n^{(1)} =& E_n^{(1)} (A \psi_n^{(1)})
\label{eq:H1_H2}
\end{eqnarray}
which serves to establish the relations between the eigenvalues, $E^{(1)}$ \& $E^{(2)}$, and eigenfunctions, $\psi^{(1)}$ \& $\psi^{(2)}$, of the superpartner Hamiltonians $H_1$ \& $H_2$, of Eq. (\ref{eq:TISE12}).
 
\begin{figure}
  \centering
  \includegraphics[width=16cm]{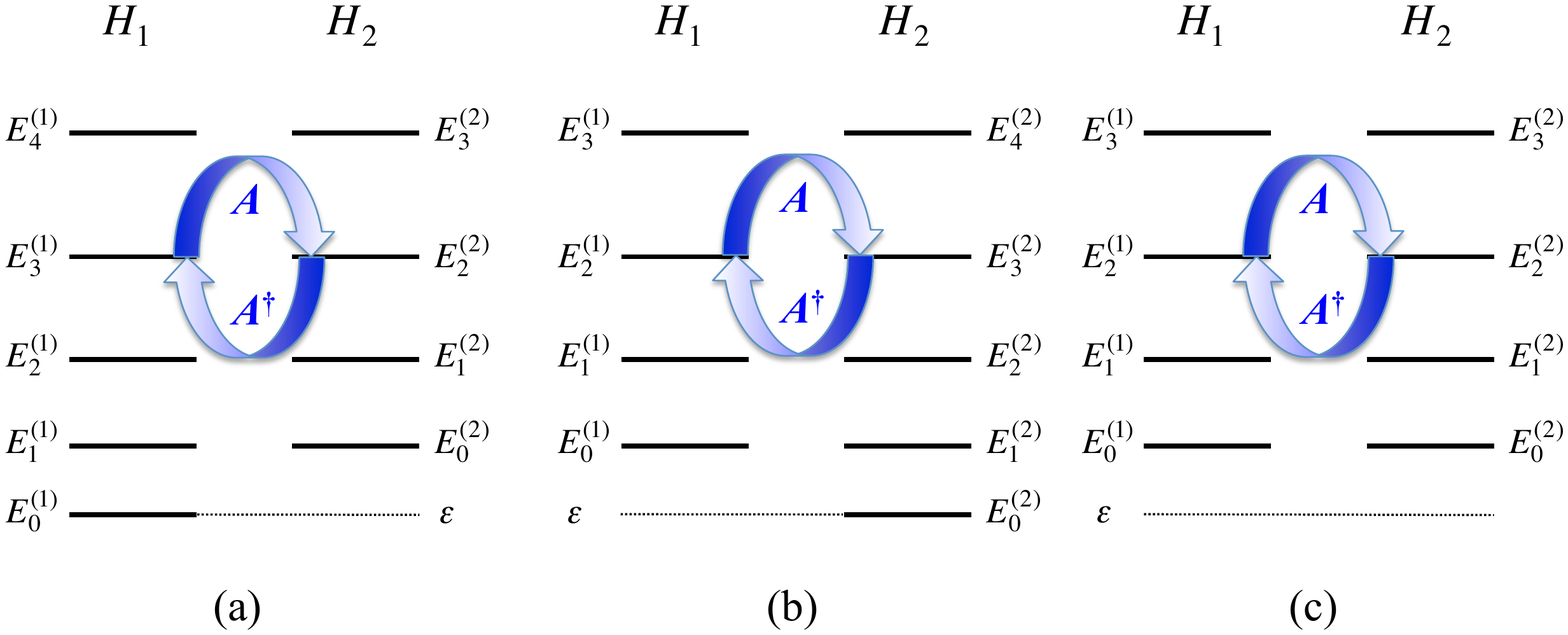}
  \caption{(Color online) Schematic diagram of the energy levels of partner Hamiltonians $H_1=A^{\dagger} A+\epsilon$ and $H_2=AA^{\dagger}+\epsilon$ for the case of standard SUSY (a), inverted SUSY (b), and broken SUSY (c). The level corresponding to the separation constant $\epsilon$ is also shown. See text. }
  \label{fig:SUSY-Scheme}
\end{figure}

For the usual choice of the constant $\epsilon$ as the ground state energy $\epsilon=E_0^{(1)}$ of Hamiltonian  $H_1$  pertaining to the wavefunction $\psi_\epsilon=\psi_0^{(1)}$, this leads to
\begin{eqnarray}
E_n^{(2)}&=& E_{n+1}^{(1)} \nonumber \\
\psi_{n  }^{(2)} &\propto&  A         \psi_{n+1}^{(1)} \nonumber \\
\psi_{n+1}^{(1)} &\propto&  A^\dagger \psi_{n  }^{(2)}
\label{eq:twins}
\end{eqnarray}
i.e., the SUSY partner Hamiltonians are isospectral, except that the ground state energy $\epsilon=E_0^{(1)}$ of $H_1$ is missing in the spectrum of $H_2$. In what follows, this will be referred to as {\it standard SUSY}, cf. Fig. \ref{fig:SUSY-Scheme}a. The intertwining operators $A$ (or $A^\dagger$) convert the eigenfunctions of $H_1$ (or $H_2$) into those of $H_2$ (or $H_1$), at the same time lowering (or raising) the respective quantum numbers by one; only the ground state eigenfunction of $H_1$ lacks a partner state but is annihilated by the intertwining operator, $A\psi_0^{(1)}=0$.

For applications to SUSY QM in position representation, the standard choice of the intertwining operators is 
\begin{eqnarray}
A(\theta)         &=&+\frac{d}{d\theta}+W(\theta) \nonumber \\
A^\dagger(\theta) &=&-\frac{d}{d\theta}+W(\theta)
\label{eq:A}
\end{eqnarray}
which leads to the following expressions for superpartner Hamiltonians
\begin{eqnarray}
H_1(\theta)&=&-\frac{d^2}{d\theta^2}+V_1(\theta) \nonumber \\
H_2(\theta)&=&-\frac{d^2}{d\theta^2}+V_2(\theta)
\label{eq:H12}
\end{eqnarray}
where the supersymmetric partner potentials $V_1(\theta)$ and $V_2(\theta)$ are related to the superpotential $W(\theta)$ via Riccati-type equations
\begin{eqnarray}
V_1(\theta)&=&W^2(\theta) - \frac{d}{d\theta}W(\theta) + \epsilon \nonumber \\
V_2(\theta)&=&W^2(\theta) + \frac{d}{d\theta}W(\theta) + \epsilon
\label{eq:riccati}
\end{eqnarray}
For a nodeless ground state wavefunction, $\psi_0^{(1)}(\theta)$, this allows to directly calculate the superpotential from a known ground state wavefunction
\begin{equation}
W(\theta)=-\frac{\frac{d}{d\theta}\psi_0^{(1)}(\theta)}{\psi_0^{(1)}(\theta)}
\end{equation}
which can be inverted to obtain an analytic expression for the wavefunction provided the superpotential is  known 
\begin{equation}
\psi_\epsilon(\theta) \propto \exp \left( -\int_0^\theta W(y)dy \right) 
\label{eq:psi}
\end{equation}
While this yields nonsingular wavefunctions for the standard choice of the ground state, $\epsilon=E_0^{(1)}$, singularities may be encountered when choosing a lower energy, $\epsilon<E_0^{(1)}$ see also our examples in Sec. \ref{sec:examples}.
In those cases, the solution $\psi_\epsilon(\theta)$ of the corresponding Schr\"{o}dinger equation, $H_1\psi_\epsilon(\theta)=\epsilon\psi_\epsilon(\theta)$, is not normalizable, even though it is still a solution of  $A(\theta)\psi_\epsilon(\theta)=0$. However, if the function
\begin{equation}
{\tilde \psi}_\epsilon(\theta)=\frac{1}{\psi_\epsilon(\theta)}
\label{eq:psi-tilde}
\end{equation}
is normalizable, it represents a ground state wavefunction ${\tilde \psi_\epsilon(\theta)}=\psi(\theta)_0^{(2)}$ of Hamiltonian $H_2$, pertaining to ground state energy $\epsilon=E_0^{(2)}$, with all other eigenstates of $H_2$ being degenerate with those of $H_1$, see Sec. 5 of Ref. \cite{s26}. In what follows, this case will be referred to as {\it inverted SUSY}, cf. Fig. \ref{fig:SUSY-Scheme}b.
If, however, ${\tilde \psi_\epsilon}$ is not normalizable and, in addition, $(d^2/d\theta^2) \log \psi_\epsilon$ is well-behaved (i.e. divergence-free and finite at the boundaries), then neither $H_1$ nor $H_2$ have a normalizable eigenstate at energy $\epsilon$, in which case $H_1$ and $H_2$ have identical spectra. 
In Sec. \ref{sec:examples} we will encounter examples both with $\epsilon=E_0^{(1)}$ and $\epsilon<E_0^{(1)}$, in the latter case with both normalizable and non-normalizable ${\tilde \psi_\epsilon}$. This case will be referred to as {\it broken SUSY}, cf. Fig. \ref{fig:SUSY-Scheme}c. 

We note in passing  that yet another case emerges by choosing $\epsilon$ to be an excited state energy, $\epsilon=E_n^{(1)}$ with
$n>0$, which gives rise to a partial or complete breakdown of the degeneracy of the energy levels of $H_1$ and $H_2$ \cite{s17}, as discussed, e.g., in our study of the quantum planar pendulum subject to combined fields \cite{SchmiFri2014b}.

\subsection{\textbf{\textit {Ansatz}} for the superpotential}
\label{sec:ansatz}

Throughout what follows we make use of the following {\it Ansatz} for the superpotential 
\begin{equation}
W(\theta) = \alpha\cot\theta+\beta\sin\theta+\gamma\csc\theta
\label{eq:W}
\end{equation}
which we introduced in Ref. \cite{SchmiFri2014b} for the case of the planar pendulum and which is an extension ($\gamma$ term added) of the {\it Ansatz} employed in Refs. \cite{LemMusKaisFriPRA2011,LemMusKaisFriNJP2011} for the case of the spherical pendulum. 
Note that the $\alpha$ term alone is related to the Rosen-Morse I superpotential whereas a combination of the $\alpha$ and $\gamma$ terms bears similarity with the P\"{o}schl-Teller I superpotential, cf. Ref. \cite{s17}.

Eq. (\ref{eq:W}) yields the following expressions for the SUSY partner potentials
\begin{eqnarray}
W^2(\theta) \mp W'(\theta) 
&=& (\alpha^2+\gamma^2\pm \alpha)\csc^2\theta \nonumber \\
&+& (2\alpha\gamma\pm \gamma)\cot\theta\csc\theta \nonumber \\
&-& (\pm \beta-2\alpha\beta)\cos\theta \nonumber \\
&-& \beta^2\cos^2\theta \nonumber \\
&-& (\alpha^2-\beta^2-2\beta\gamma)
\label{eq:V12}
\end{eqnarray}
By identifying the effective potential (\ref{eq:pot}) for the quantum rotor subject to the combined interactions with $V_1 =W^2-W'+\epsilon$, we obtain:
\begin{eqnarray}
m^2 - \tfrac{1}{4} &=& \alpha^2+\gamma^2 + \alpha \nonumber \\
0                  &=& 2\alpha\gamma + \gamma   \nonumber \\
\eta               &=& \beta(1-2\alpha) \nonumber \\
\zeta              &=& \beta^2 \nonumber \\
\epsilon           &=& \alpha^2-\beta^2-2\beta\gamma -\tfrac{1}{4}
\label{eq:eze}
\end{eqnarray}
In order for the first two equalities to hold, one of the following conditions must be fulfilled: 
\begin{eqnarray}
&\text{Case 1:}& \alpha=\pm m-\tfrac{1}{2} \,{\text {and}}\, \gamma=0 \\ 
&\text{Case 2:}& \alpha= -\tfrac{1}{2} \,{\text {and}}\, \gamma=\pm m.
\end{eqnarray} 
with $m\equiv |m|$. Where necessary, we will subdivide each of these two cases into two subcases referred to as $1_+$, $1_-$, $2_+$, $2_-$, depending on the sign of $m$.
Below, we will discuss these cases individually and show that they are connected with a particular ratio of the orientation to alignment parameters and, therefore, with particular intersections of the eigenenergy surfaces characterized in Sec. \ref{sec:combined}, namely Case $1_{\pm}$ with $k=\frac{\eta}{2\sqrt{\zeta}}=\mp m+1$ and Case 2 with $k=\frac{\eta}{2\sqrt{\zeta}}=1$. Thereby we are substantially extending the work in Refs. \cite{LemMusKaisFriPRA2011,LemMusKaisFriNJP2011} where only Case $1_-$ was considered. Obviously, for $m=0$ the two cases coincide yielding $k=1$, a SUSY and eigensurface topology discussed in our previous work on the spherical pendulum in combined fields \cite{SchmiFri2014a}.
It is worth noting that for a hypothetical value of $m=\frac{1}{2}$, Cases $1_+$, $2_\pm$, $1_-$ yield $\kappa=2k=1,2,3$, which was discussed in our work on the SUSY and eigensurface topology of the planar pendulum \cite{SchmiFri2014b}.

The knowledge of the superpotential $W$ makes it possible to construct the supersymmetric partner potential $V_2 =W^2+ W'+\epsilon$, which -- apart from changes in the singular terms proportional to $\csc \theta$ or $\cot \theta$ -- differs from $V_1$ in that the interaction parameter $\eta$ is effectively altered, see Eq. (\ref{eq:V12}) and (\ref{eq:eze}). 
Furthermore, using Eq. (\ref{eq:psi}) one can derive an analytic expression   
for the wavefunction  from the superpotential $W$ pertaining to the energy eigenvalue $\epsilon$ obtained from the last line of Eq. (\ref{eq:eze}).
For the particular superpotential introduced by Eq. (\ref{eq:W}), the wavefunction takes the form
\begin{eqnarray}
\psi_\epsilon(\theta) 
&\propto& (\csc\theta)^\alpha\,\exp(\beta\cos\theta)\,\left(\cot\frac{\theta}{2}\right)^\gamma\nonumber \\ 
&\propto& \left(\csc\frac{\theta}{2}\right)^{\alpha+\gamma}\,\left(\sec\frac{\theta}{2}\right)^{\alpha-\gamma}\,\exp(\beta\cos\theta)
\label{eq:psi_abc}
\end{eqnarray}

We note that identifying potential (\ref{eq:pot}) with $V _2=W^2+W'+\epsilon$ instead of $V_1 $ would only lead to an interchange of the roles of $V _1$ and $V _2$, combined with a sign change of $m$. Thus such an interchange would not furnish any new superpotential or analytic wavefunction.

\subsection{Case 1: Supersymmetric solutions with {\boldmath $k=\mp m+1$}}
\label{sec:susy_1}

For $\alpha=\pm m-\tfrac{1}{2},\,\gamma=0$ the superpotential (\ref{eq:W}) simplifies to
\begin{equation}
W=\left(\pm m-\tfrac{1}{2}\right)\cot\theta+\beta\sin\theta
\label{eq:W_1}
\end{equation}
and Eq. (\ref{eq:eze}) yields the following expressions for the interaction parameters and the energy in terms of the parameter $\beta$ of Eq. (\ref{eq:W}),
\begin{eqnarray}
\eta     &=& 2\beta(\mp m+1) \nonumber \\
\zeta    &=& \beta^2 \nonumber \\
\epsilon &=& m(m \mp 1)-\beta^2
\label{eq:eze_1}
\end{eqnarray}
A comparison of the above expressions for the interaction parameters with Eq. (\ref{eq:k}) reveals that the Case 1 solutions correspond to the loci of the intersections with topological index $k=\mp m+1$. Hence, for Case 1 the value of the topological index becomes a function of the azimuthal quantum number $m$, see also examples in Sec. \ref{sec:examples}.
  
The Case 1 SUSY partner potentials, obtained from Eq. (\ref{eq:eze}), become
\begin{eqnarray}
V_1 (\theta) &=& \left(m^2-\tfrac{1}{4}\right)\csc^2\theta - 2\beta(\mp m + 1)\cos\theta - \beta^2\cos^2\theta -\tfrac{1}{4} \nonumber \\
V_2 (\theta) &=& \left[(m \mp 1)^2-\tfrac{1}{4}\right]\csc^2\theta \pm 2\beta m\cos\theta - \beta^2\cos^2\theta -\tfrac{1}{4}
\label{eq:V12_1}
\end{eqnarray}
Inspection of the above partner potentials reveals that 
$V _2$ is of the same type as $V _1$ but with altered azimuthal quantum number, ${\tilde m}$, in the $\csc^2\theta$--term: in Case $1_+$, $\tilde{m}=m-1$, whereas in Case $1_-$, $\tilde{m}=m+1$.
Given that, in Case 1, $k=\mp m+1$, we see that the topological index appearing in the $\cos\theta$--term of $V _2$ is ${\tilde k}=k-1$. 
In other words, $V _2$  differs from $V _1$ in that the interaction parameter $\eta=2\beta k$ is effectively  reduced by $2\beta$. In addition we note  that the Case 1 $V _1$ potentials are qualitatively different  for $\pm m$, whereas the $V _2$ are just mirror images of each other for $\pm m$. 
See also Sec. \ref{sec:examples} where the interrelations between SUSY partners of the combined field rotor Hamiltonians with different $m$ and $k$ are discussed in some detail.

From Eq. (\ref{eq:psi_abc}) we obtain the nodeless Case 1 eigenfunction of the original Hamiltonian $H_1$
\begin{equation}
\psi_{\epsilon}(\theta) \propto (\sin\theta)^{\mp m + \frac{1}{2}} \exp (\beta\cos\theta)
\label{eq:psi_1}
\end{equation}
For Case 1$_-$, this solution is normalizable and represents the  ground state $\psi_\epsilon=\psi_0^{(1)}$ of $H_1$ with zero point energy $\epsilon=E_0^{(1)}$, i.e., we encounter here standard SUSY with $H_2$  isospectral with $H_1$ but having one bound state less. However, for Case 1$_+$ and for $m>0$, this solution becomes singular for $\theta\rightarrow 0$ with  $\epsilon<E_0^{(1)}$. As mentioned above, the reciprocal wavefunction 
${\tilde \psi}_\epsilon=1/\psi_\epsilon$ represents a normalizable  ground state ${\tilde\psi}_\epsilon=\psi_0^{(2)}$ of $H_2$ with zero point energy $\epsilon=E_0^{(2)}$, i.e., we encounter inverted SUSY, with $H_2$  isospectral with $H_1$ but having one bound state more. See also our examples in Sec. \ref{sec:examples} as well as Fig. \ref{fig:cut3}.

\subsection {Case 2: Supersymmetric solutions with {\boldmath $k=1$}}
\label{sec:susy_2}

For $\alpha=-\tfrac{1}{2},\,\gamma=\pm m$ the superpotential (\ref{eq:W}) becomes
\begin{equation}
W=-\tfrac{1}{2}\cot\theta+\beta\sin\theta \pm m\csc\theta
\label{eq:W_2}
\end{equation}
and Eq. (\ref{eq:eze}) yields the following expressions for the interaction strength parameters and for the energy in terms of $\beta$,\begin{eqnarray}
\eta&=&2\beta \nonumber \\
\zeta&=&\beta^2 \nonumber \\
\epsilon &=& \mp 2\beta m -\beta^2 
\label{eq:eze_2}
\end{eqnarray}
A comparison of the above expressions for the interaction parameters with Eq. (\ref{eq:k}) reveals that the Case 2 solutions correspond to the loci of the intersections with topological index $k=1$.
Hence, in Case 2 the value of the topological index is independent of the azimuthal quantum number $m$, see also examples in Sec. \ref{sec:examples}.

The corresponding SUSY partner potentials for Case 2 take the form
\begin{eqnarray}
V_1 (\theta)&=& \left(m^2-\tfrac{1}{4}\right)\csc^2\theta - 2\beta\cos\theta - \beta^2\cos^2\theta -\tfrac{1}{4} \nonumber   \\
V_2 (\theta)&=& \left(m^2+\tfrac{3}{4}\right)\csc^2\theta \mp 2 m\cot\theta \csc\theta - \beta^2\cos^2\theta -\tfrac{1}{4}
\label{eq:V12_2}
\end{eqnarray}
Thus the Case 2 $V _1$ potentials are the same for $\pm m$ whereas the $V _2$ potentials are mirror images of each other  for $\pm m$.
Note that the partner potentials $V_1$ and $V_2$ are not of the same type since the prefactor of the $\csc^2\theta$ in $V_2$ cannot be rewritten as ${\tilde m}^2-1/4$ (with integer $\tilde{m}$) and because of an additional term in $V_2$ for $m\ne 0$ proportional to $\cot\theta\csc\theta$.

From Eq. (\ref{eq:psi_abc}) we obtain the nodeless Case 2 eigenfunctions for the original potential $V_1$
\begin{equation}
\psi_\epsilon(\theta) \propto (\sin\theta)^{\frac{1}{2}} (\tan\tfrac{\theta}{2})^{\mp m}\exp (\beta\cos\theta)
\label{eq:psi_2}
\end{equation}
For $m=0$ this solution is normalizable and represents the $H_1$ ground state $\psi_\epsilon=\psi_0^{(1)}$  with zero point energy $\epsilon=E_0^{(1)}$. 
For Case $2_+$ with $m>0$, however, this solution becomes singular for $\theta\rightarrow 0$ with  $\epsilon<E_0^{(1)}$.  Also the reciprocal wavefunction 
${\tilde \psi}_\epsilon=1/\psi_\epsilon$ exhibits a singularity (for $\theta\rightarrow \pi$) so that neither $\psi_\epsilon$ nor ${\tilde \psi}_\epsilon$ represent normalizable ground state wavefunctions for $H_1$ or $H_2$, see also our examples in Sec. \ref{sec:examples}. This is broken SUSY where $H_1$ and $H_2$ are exactly isospectral (i.e., exhibit one-to-one pairing of eigenstates). The same is true for Case $2_-$, even though both $\psi_{\epsilon}$ and $\psi^{-1}_{\epsilon}$ are normalizable, see also Fig. \ref{fig:cut3}.

\section{Examples}
\label{sec:examples}
The eigenproperties below were calculated with the Fourier Grid Hamiltonian (FGH) method \cite{Meyer:70a,Marston:89a} as implemented in WavePacket software \cite{Schmidt-WavePacket4.9} with 512 grid points. Energies above 10000 have been truncated. 

\subsection{$m=0$}
\label{sec:examples_m0}
\begin{figure}
  \centering
  \includegraphics[width=14cm]{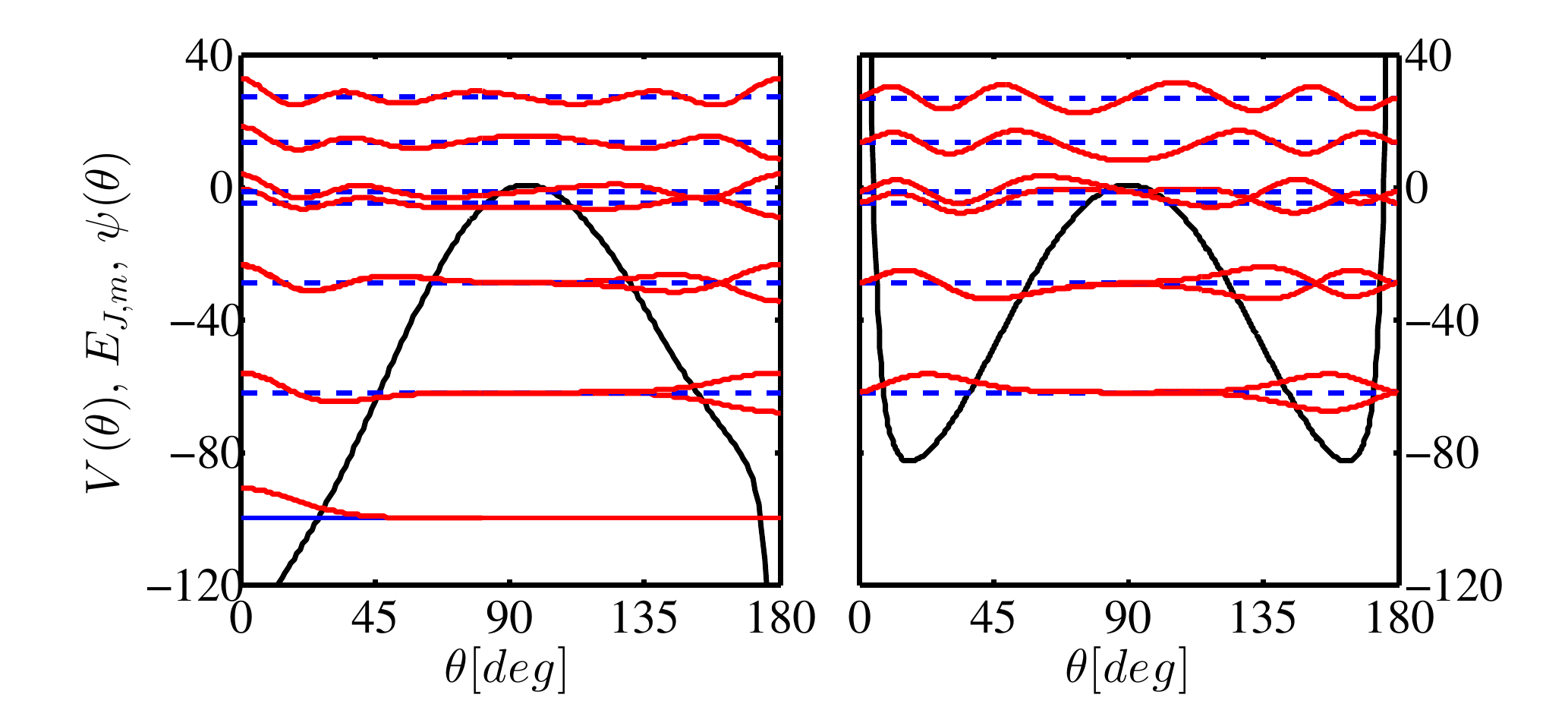}
  \caption{(Color online) Cases $1$ and $2$ for $m=0$: Superpartner potentials $V_1$ (left) corresponding to  $k=1$ and $V_2^+$ (right) corresponding to $\tilde{m}=-1$ and $\tilde{k}=0$, eigenenergies (blue lines) and eigenfunctions (red curves) of Hamiltonian (\ref{eq:hamilton}) for $\beta=10$. Analytic eigenenergies are shown by full blue lines.}
  \label{fig:m=0_k+1}
\end{figure}

Here the partner potential $V_1 (\theta)$ as well $V_2 (\theta)$ is the same in Cases 1 and 2, cf. Eqs. (\ref{eq:V12_1}) and (\ref{eq:V12_2}) and Table \ref{tab:partners}.  Figure \ref{fig:m=0_k+1} shows the superpartner potentials, eigenenergies and eigenfunctions for $\beta=10$. Note that  the ground-state energy $\epsilon=E_0^{(1)}=-\beta^2=-100$ and wavefunction 
\begin{equation}
\psi^{(1)}_{0}(\theta) \propto (\sin\theta)^{ \frac{1}{2}} \exp (\beta\cos\theta)
\label{eq:psi_1_0}
\end{equation}
are analytic. As discussed in Sec. \ref{sec:susy_1}, here $k=1$, $m=1$ for $H_1$  change to $\tilde{k}=0$, $\tilde{m}=1$ for $H_2$. Hence the superpartner $V_2 (\theta)$ is a  symmetric double-well potential corresponding to the purely aligning interaction which leads to the oblate spheroidal wave equation, see Sec. \ref{sec:zeta}. However, the analytically available eigenenergies lie below its potential's minimum, which is the standard SUSY case where $H_2$ has one state less than $H_1$. In 3D, the wavefunction $\chi(\theta)\propto \exp (\beta\cos\theta)$, cf. Eq. (\ref{chi}), has the form of the von Mises function \cite{Abramovitz:72a}, which can be viewed as a generalization of a Gaussian distribution for the case of a random variable defined on a circle. We note that the von Mises distribution comes out only for combined fields where both $\eta$ and $\zeta$ are nonzero.

\subsection{$m=1$}
\label{sec:examples_m1}
\begin{figure}
  \centering
  \includegraphics[width=14cm]{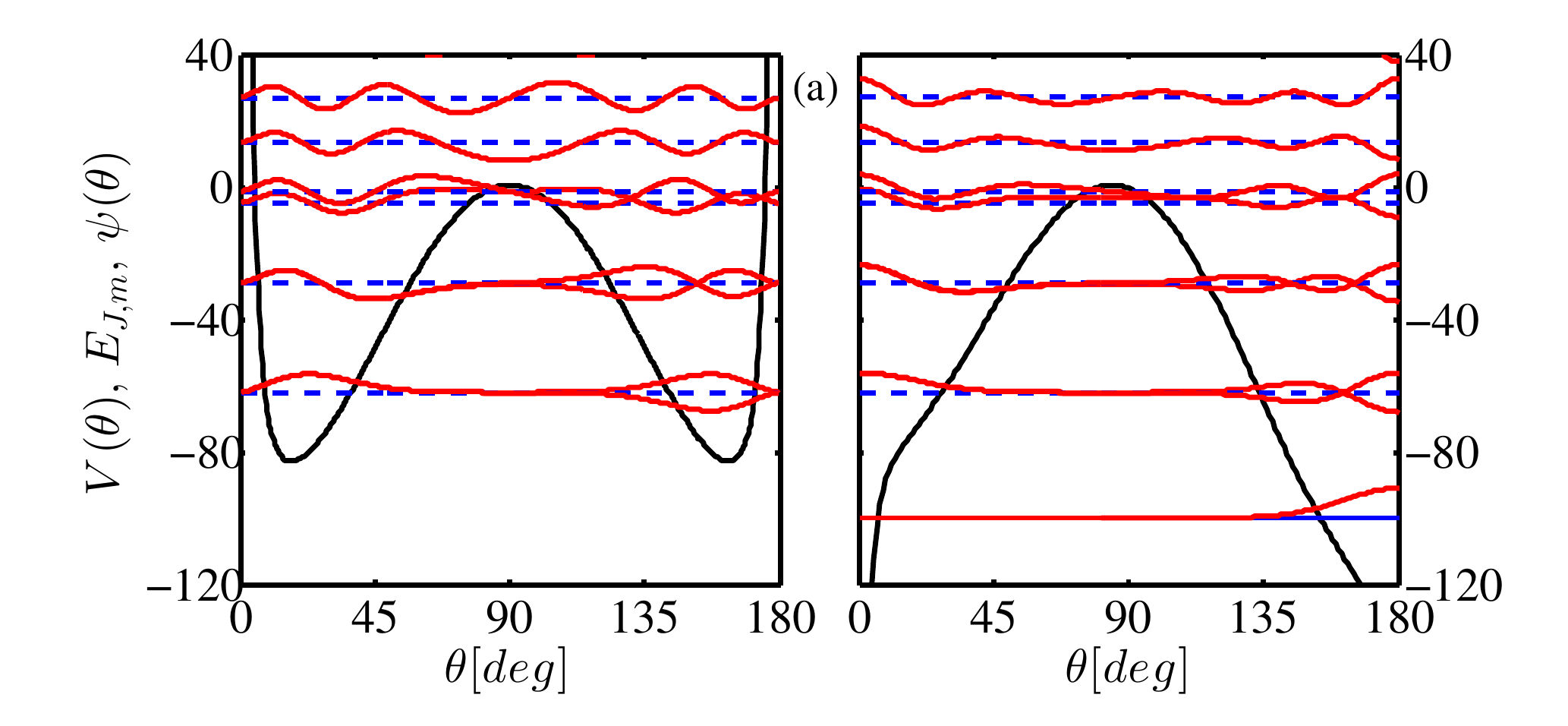}
  \includegraphics[width=14cm]{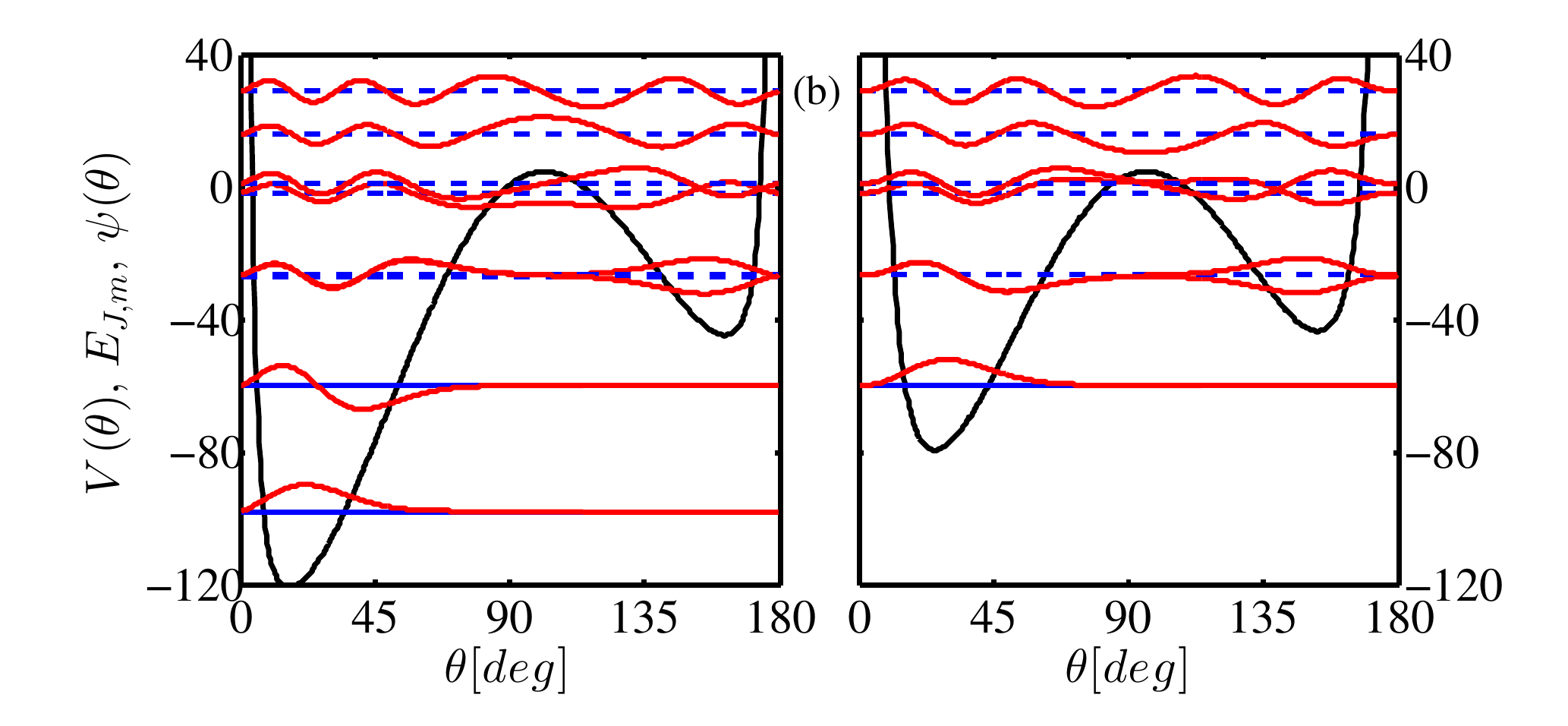}
  \includegraphics[width=14cm]{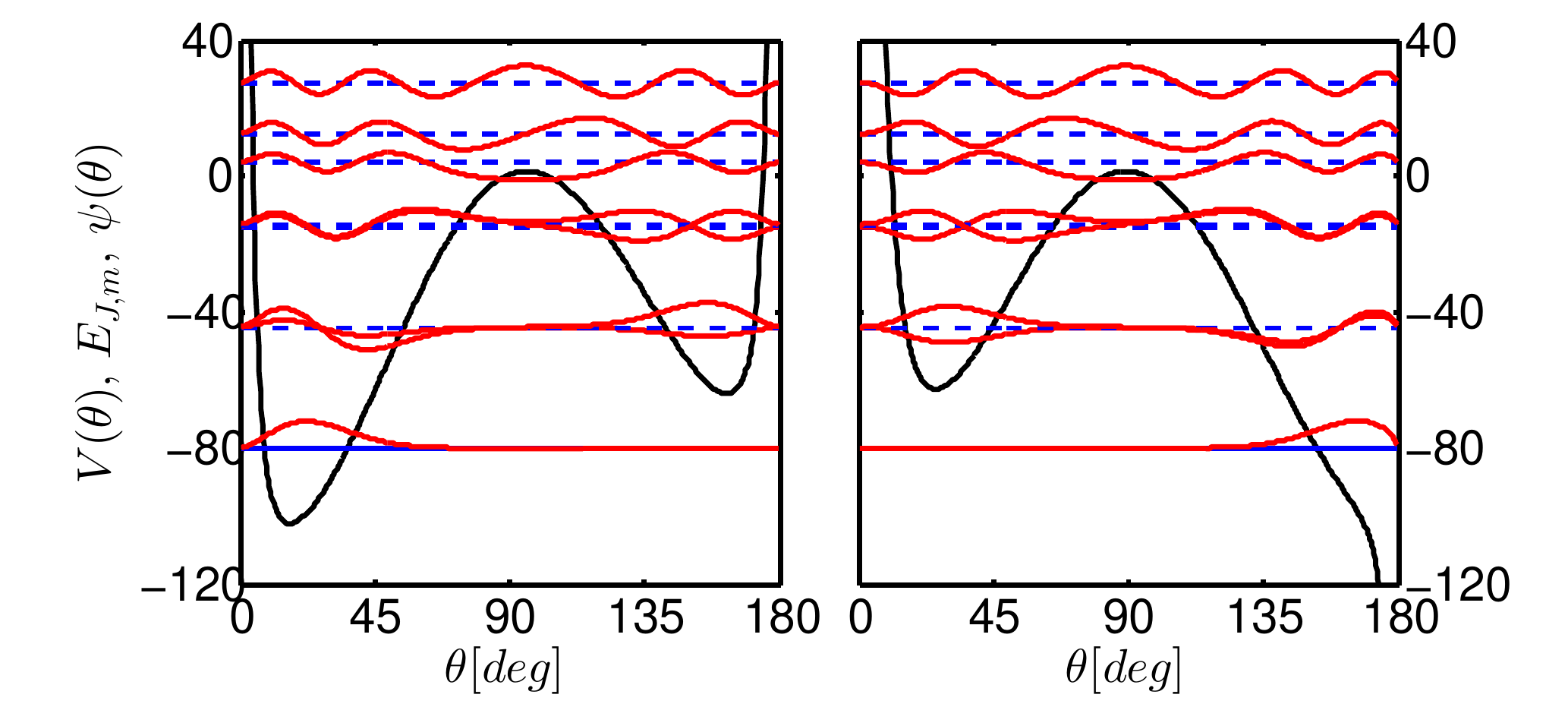}
  \caption{(Color online) Cases 1$_+$ (a), 1$_-$ (b), and 2$_-$ (c) for $m=1$.  Superpartner potentials $V_1$ (left) and $V_2$ (right), eigenenenergies (blue lines) and eigenfunctions (red curves) of Hamiltonian (\ref{eq:hamilton}) for $\beta=10$. Analytic eigenenergies are shown by full blue lines.}
  \label{fig:m_1}
\end{figure}

In Case $1_+$,  $k=0$, $\tilde{m}=0$, and $\tilde{k}=-1$, cf. Fig. \ref{fig:m_1}a, drawn for $\beta=10$, and Table \ref{tab:partners}. The symmetric double well potential $V_1 $ yields an eigenenergy $\epsilon=-\beta^2=-100$, which lies below the minimum of $V_1 (\theta)$ and $\psi_{\epsilon}\propto (\sin\theta)^{-\frac{1}{2}}\exp(\beta\cos\theta)$ is not normalizable. However,  the reciprocal, $\tilde{\psi}_{\epsilon}=1/\psi_{\epsilon}$, is normalizable. Thus, whereas $\psi_{\epsilon}$ is not a meaningful state of the original Hamiltonian $H_1$, its reciprocal $\tilde{\psi}_{\epsilon}$ yields the ground state wavefunction of $H_2$ with $\epsilon=E_0^{(2)}$. Hence $H_1$ and $H_2$ are isospectral, except  that $H_1$ has one state less than $H_2$, which is characteristic for the inverted SUSY scenario. 

A comparison of these partner potentials and eigenproperties with the above $m=0$ case reveals that the two cases are similar, but that the roles of the partner potentials have been interchanged, $V_1 \leftrightarrow V_2 $ and that the asymmetry of the $V_2 $ potential for $m=1$ has been reversed with respect to that of the $V_1 $ potential for $m=0$, $\theta \rightarrow \pi -\theta $. As a result, $\epsilon=E_0^{(2)}=-\beta^2=-100$ and $\psi_0^{(2)}$ is normalizable and  equal to $\psi_0^{(1)}(\pi-\theta)$ for  $m=0$.

In Case $1_-$,  $k=2$, $\tilde{m}=2$, and $\tilde{k}=1$, cf. Fig. \ref{fig:m_1}b, drawn for $\beta=10$, and Table \ref{tab:partners}. As a result, the eigenenergy $\epsilon=E_0^{(1)}=2-\beta^2=-98$ and the eigenfunction $\psi_0^{(1)}\propto(\sin\theta)^{\frac{3}{2}}\exp(\beta\cos\theta)$, cf. Eq. (\ref{eq:psi_1}), is normalizable. Hence, we have the standard SUSY case where $H_1$ and $H_2$ are isospectral except that $H_2$ has one state less (because $\psi_0^{(1)}$ does not have a partner state). 

In Case $2_+$, $k=1$, cf. Table \ref{tab:partners}, the eigenenergy  $\epsilon=-2\beta-\beta^2=-120$ (calculated for $\beta=10$) lies far below the minimum of the $V_1 $ potential and neither $\psi_{\epsilon}$ nor $\tilde{\psi}_{\epsilon}$ are normalizable. Hence, we have the case of broken SUSY, the characteristic feature of which is that $H_1$ and $H_2$ are exactly isospectral, with a one-to-one pairing of eigenstates.

This feature can be also seen in the closely related Case $2_-$,  $k=1$,  cf. Fig. \ref{fig:m_1}c, drawn for $\beta=10$, and Table \ref{tab:partners}.  There, however, the eigenenergy $\epsilon=E_0^{(1)}=2\beta-\beta^2=-80$ lies within both superpartner potentials and the corresponding wavefunction, $\psi_0^{(1)}\propto(\sin\theta)^{\frac{1}{2}}\tan(\frac{\theta}{2})\exp(\beta\cos\theta)$, cf. Eq. (\ref{eq:psi_2}), and its reciprocal are both normalizable. Note that also here  the partner Hamiltonians $H_1$ and $H_2$ are  exactly isospectral, which indicates broken SUSY for both Cases $2_+$ and $2_-$.

\subsection{$m=2$}
\label{sec:examples_m2}

\begin{figure}
  \centering
  \includegraphics[width=14cm]{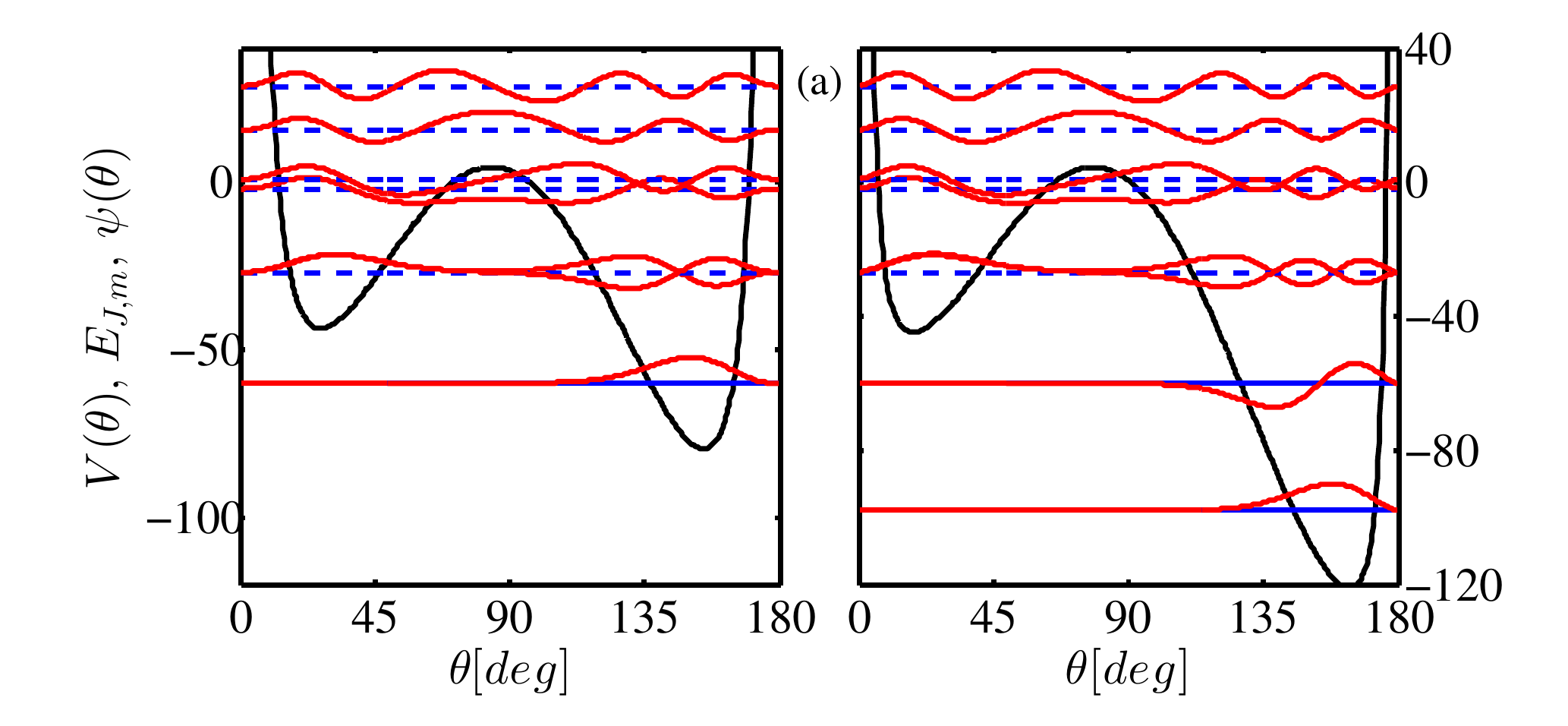}
  \includegraphics[width=14cm]{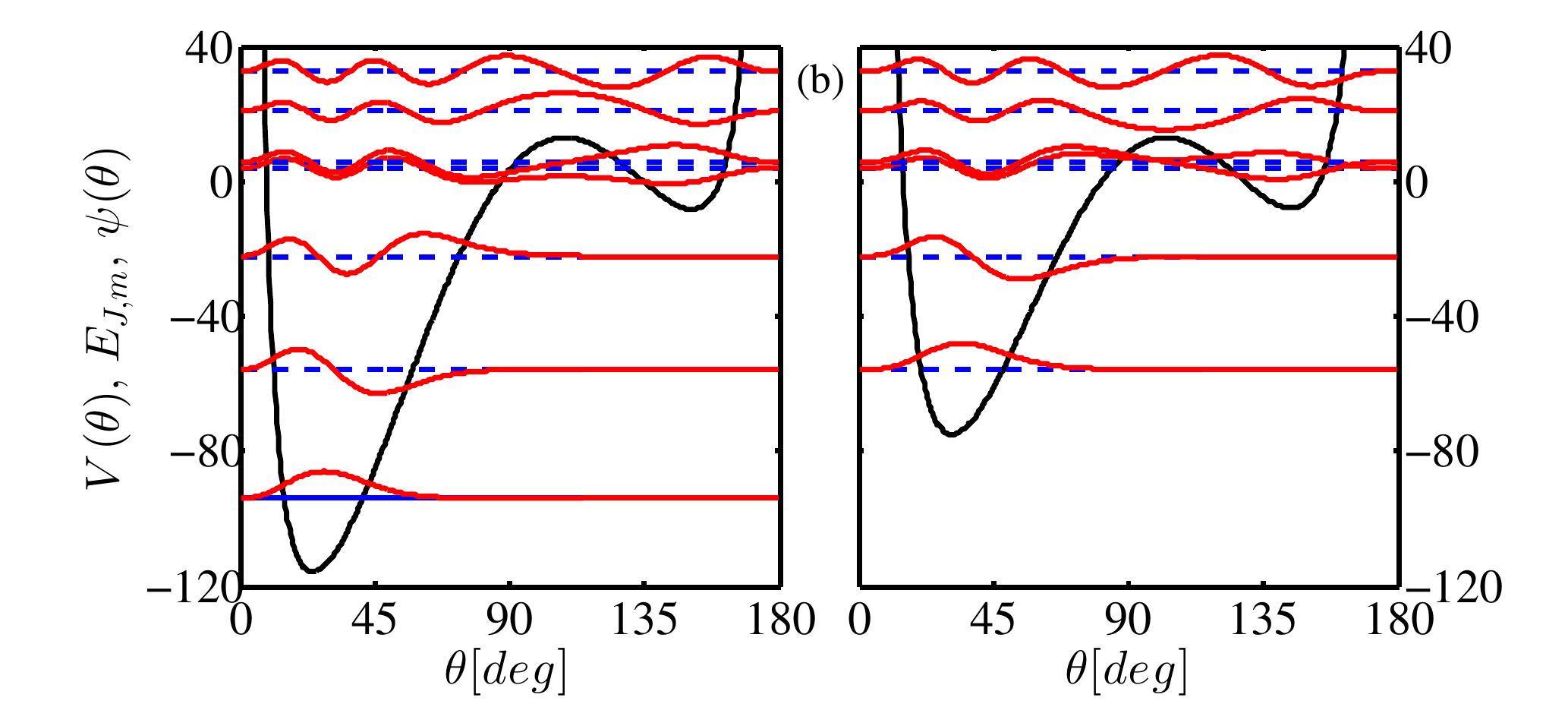}
  \includegraphics[width=14cm]{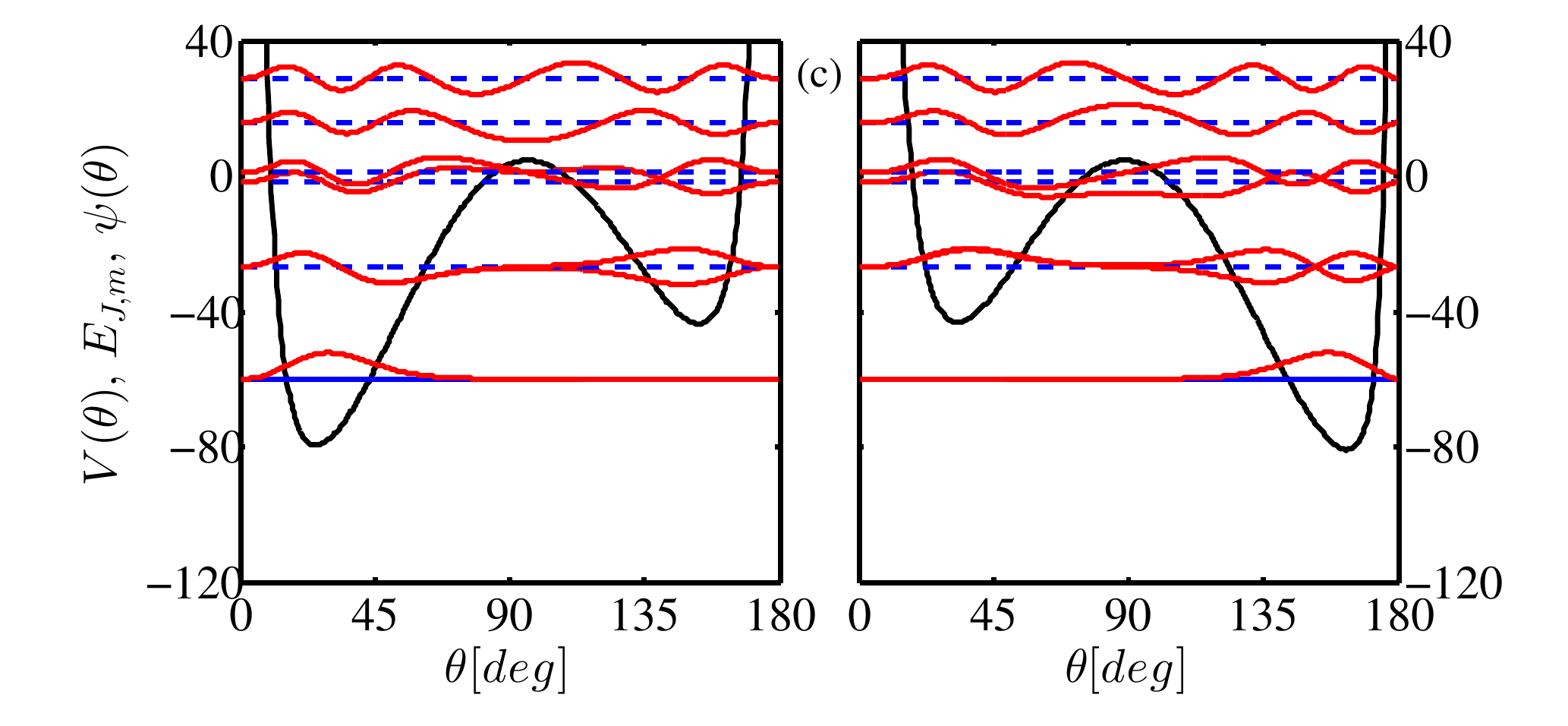}
  \caption{(Color online) Cases 1$_+$ (a), 1$_-$ (b), and 2$_-$ (c) for $m=2$.  Superpartner potentials $V_1$ (left) and $V_2$ (right), eigenenenergies (blue lines) and eigenfunctions (red curves) of Hamiltonian (\ref{eq:hamilton}) for $\beta=10$. Analytic eigenenergies are shown by full blue lines.}
  \label{fig:m_2}
\end{figure}

In Case $1_+$,  $k=-1$, $\tilde{m}=1$, and $\tilde{k}=-2$, cf. Fig. \ref{fig:m_2}a,  drawn for $\beta=10$, and Table \ref{tab:partners}. Here we have again the inverted SUSY case with $\epsilon=2-\beta^2=-98$ for the example shown. Hence whereas $\psi_{\epsilon}$ is not a meaningful state of the original Hamiltonian $H_1$, its reciprocal, $\tilde{\psi}_{\epsilon}$, yields the ground state wavefunction for $H_2$ with $\epsilon=E_0^{(2)}$. Hence $H_1$ and $H_2$ are isospectral, except that $H_1$ has one state less than $H_2$, which is characteristic of inverted SUSY. Note that $V_2$ is a mirror image of $V_1$ for Case $1_-$ with $m=1$ and $k=2$ (discussed above in Sec. \ref{sec:examples_m1}), cf. Fig. \ref{fig:m_1}b. 

In Case $1_-$,  $k=3$, $\tilde{m}=3$, and $\tilde{k}=2$, cf. Fig. \ref{fig:m_2}b, drawn for $\beta=10$, and Table \ref{tab:partners}. Hence, we have the case of standard SUSY, with $\epsilon=E_0^{(1)}=6\beta-\beta^2=-94$ where $H_1$ and $H_2$ are isospectral except that $H_2$ has one state less (because $\psi_0^{(1)}$ does not have a partner state).

In Case $2_+$, $k=1$, cf. Table \ref{tab:partners}. Here $\epsilon=-4\beta-\beta^2=-140$ (for our example with $\beta=10$) and neither $\psi_{\epsilon}$ nor $\tilde{\psi}_{\epsilon}$ is normalizable.
In Case $2_-$, $k=1$, cf. Fig. \ref{fig:m_2}c, drawn for $\beta=10$, and Table \ref{tab:partners}. Here $\epsilon=E_0^{(1)}=E_0^{(2)}=4\beta-\beta^2=-60$ and both $\psi_{\epsilon}\propto(\sin\theta)^{\frac{1}{2}} (\tan\frac{\theta}{2})^2\exp(\beta\cos\theta)$ and $\tilde{\psi}_{\epsilon}$ are normalizable. In both these cases we encounter broken SUSY, which implies exactly isospectral $H_1$ and $H_2$.

We note that the Case $2_-$ Hamiltonian for $m=2$, $k=1$ can be also regarded as a Case $1_-$ superpartner $H_2$ with $\tilde{m}=2$ and $\tilde{k}=1$ of the original Hamiltonian $H_1$ for $m=1$, $k=2$, as we have already discussed in Sec. \ref{sec:examples_m1}.
Hence, the ground state of the  Hamiltonian $H_2$ with $\tilde{m}=2$, $\tilde{k}=1$,
\begin{equation}
    \psi_0^{(2)} \propto (\sin\theta)^{\frac{1}{2}} (\tan \tfrac{\theta}{2})^2 \exp (\beta\cos\theta)
\end{equation}
with energy $E_0^{(2)}=4\beta - \beta^2=-60$, cf. Fig. \ref{fig:m_2}c, can be used to construct an isoenergetic wavefunction for the first excited state of the Hamiltonian $H_1$ with $m=1$, $k=2$ by virtue of the intertwining relation, Eq. (\ref{eq:twins}), for the standard SUSY case
\begin{equation}
    \psi_1^{(1)} \propto A^\dagger \psi_0^{(2)} \propto
    \exp(\beta\cos\theta) (\cos\theta-1) (\beta\cos\theta-\beta+1)(\sin\theta)^{-\frac{1}{2}}
\end{equation}
where the intertwining operator $A^\dagger=-\frac{d}{d\theta}+W$ is based on the superpotential $W=-\frac{3}{2}\cot\theta+\beta\sin\theta$ obtained from Eq. (\ref{eq:W_1}) for $m=1$, Case $1_-$.
Hence, the Hamiltonian $H_1$ for $m=1$, Case $1_-$ ($k=2$) is  the only case, for which our SUSY analysis has rendered  analytic expressions for both the ground and the first excited state eigenenergies and wavefunctions.
Note that similar findings were made for the planar quantum pendulum \cite{SchmiFri2014b}, where two analytic wavefunctions could be obtained for one value of the topological index (there $\kappa=2$).

\subsection{Free rotor limit}

In the field-free limit, $\eta,\zeta\rightarrow 0$, Hamiltonian (\ref{eq:hamilton}) reduces to that of a rigid rotor, $\mathcal{H}=\mathbf{J}^2$. Although its analytic eigenenergies, $E_{j,m}=j(j+1)$, and eigenfunctions, $\psi_{j,m}\propto Y_{j,m}$, are well-known, it is instructive to have a look at the free rotor from the SUSY QM perspective.  

Our starting point is Case $1_-$, cf. Sec. \ref{sec:susy_1}, as this standard SUSY case is especially straightforward to adapt. In the free rotor limit, $\beta\rightarrow 0$, the  superpotential of Eq. (\ref{eq:W_1}) reduces to 
\begin{equation}
W(\theta)=-(m+\tfrac{1}{2})\cot\theta
\label{eq:W_free}
\end{equation}
and so the ground-state wavefunction (\ref{eq:psi_1}) simplifies to 
\begin{eqnarray}
\psi_0^{(1)}(\theta)      &\propto& (\sin\theta)^{m+1/2} \nonumber \\
\chi_0^{(1)}(\theta,\phi) &\propto& (\sin\theta)^{m    } e^{im\phi}
\label{eq:stretch}
\end{eqnarray}
where Eq. (\ref{chi}) has been used to convert the  wavefunctions $\psi(\theta)$ pertaining to the effective 1D potential to their 3D counterparts, $\chi(\psi,\theta)$. It can be immediately seen that the $\chi(\psi,\theta)$ wave functions are proportional to spherical harmonics, $Y_{m,m}$, 
for ``stretched states,'' i.e., rotational states with maximal projection quantum number, $m=j$. The corresponding superpartner potentials obtained from Eq. (\ref{eq:V12_1}) take the form:
\begin{eqnarray}
V_1(\theta) &=& \left( m^2-\tfrac{1}{4}\right) \csc^2\theta-\tfrac{1}{4} \nonumber \\
V_2(\theta) &=& \left( [m+1]^2-\tfrac{1}{4}\right) \csc^2\theta-\tfrac{1}{4}
\end{eqnarray}
which reveals that the superpartner Hamiltonian $H_2$ is of the same form as the original Hamiltonian $H_1$ but with $\tilde{m}=m+1$. Hence the spectrum of $H_2$ is the same as that of $H_1$ except that the ground state of $H_1$ is missing in the spectrum of $H_2$. This is, of course, expected for free rotor states with azimutal quantum number $m$ incremented by one and can be gleaned from the left borders, corresponding to $\eta\rightarrow 0$ or  $\zeta\rightarrow 0$, of the upper panels of Figs. \ref{fig:eta3} or \ref{fig:zeta3}. In an analogous way, a Hamiltonian $H_3$ can be constructed that is isospectral with $H_2$, except that the ground state of $H_3$ is missing in the spectrum of $H_2$. 
This procedure, which can be repeated indefinitely, leads to a hierarchy of isospectral Hamiltonians which are said to be ``shape invariant.'' The property of shape invariance is a sufficient condition for analytic solvability, i.e., closed-form expressions can be found for all stationary states \cite{LemMusKaisFriNJP2011}.

The analytic solvability of the free-rotor problem follows from the  relations (\ref{eq:twins}) for the intertwining operators $A$ and $A^\dagger$ defined by Eq. (\ref{eq:A}) for the superpotential $W(\theta)$ given by Eq. (\ref{eq:W_free}). The intertwining operator $A(\theta)=d/d\theta+W(\theta)$ acting on the ground state wavefunction of $H_1$, $\psi_0^{(1)}(\theta)\propto (\sin\theta)^{m+\frac{1}{2}}$, yields zero and, accordingly, the ground state of $H_1$ lacks a partner state. Conversely,  $A^\dagger(\theta)=-d/d\theta+W(\theta)$ acting on the  ground state wavefunction of $H_2$, $\psi_0^{(2)}(\theta) \propto (\sin\theta)^{m+3/2}$ (corresponding to $Y_{m+1,m+1}$), yields the first excited state wavefunction of $H_1$
\begin{eqnarray}
\psi_1^{(1)}(\theta)      &\propto& \cos\theta(\sin\theta)^{m+1/2} \nonumber \\
\chi_1^{(1)}(\theta,\phi) &\propto& \cos\theta(\sin\theta)^{m    } e^{im\phi}
\end{eqnarray}
where the latter expression corresponds to spherical harmonics $Y_{m+1,m}$. Hence, the intertwining operator $A^\dagger$ serves to lower the azimuthal quantum number $m$ without changing the quantum number $j$  of the free rotor states. In close analogy, $A$  raises the quantum number $m$, without changing $j$. These findings are not  surprising since the intertwining operators $A$ and $A^\dagger$ for the free rotor are equivalent to the rising and lowering ladder operators $J_+=J_x+iJ_y$ and  $J_-=J_x-iJ_y$, respectively \cite{Varshalovich2008}. (Note, however, that the intertwining operators connect states of different Hamiltonians.)  Thus by repeated action of  $A^\dagger$, all rotational states (i.e., all spherical harmonics) can be generated via Eq. (\ref{eq:stretch}) for the stretched-state wavefunction.

\section{Conclusions and Outlook}
\label{sec:outlook}

We undertook a mutually complementary  analytic and computational study of the full-fledged spherical (3D) quantum rotor subject to combined orienting and aligning interactions, characterized, respectively, by dimensionless parameters $\eta$ and $\zeta$. These interactions correspond to those of a polar (permanent electric dipole moment) and polarizable (induced electric dipole moment) molecule subject to either an electrostatic field or a combination of an electric and an far-off-resonant optical field or a single-cycle non-resonant optical field. 

We considered a full range of interaction strengths, which convert, jointly or separately, the spherical rotor into a hindered rotor or a quantum pendulum or a harmonic librator (angular harmonic oscillator), depending on the values of $\eta$ and $\zeta$. Following upon our previous studies of both planar and spherical pendula, we were concerned with the topology of the eigenenergy surfaces spanned by the interaction parameters $\eta$ and $\zeta$ for all values of the projection quantum number $m$ as well as with the supersymmetry of the eigenproblem as a means for identifying its analytic solutions.

{\it Topology}. We found that the loci of all  intersections that arise among the eigenenergy surfaces of the  quantum pendulum are accurately rendered by a simple formula, Eq. (\ref{eq:k}). Furthermore, since the equation for the loci is independent of the eigenstate, the energy levels exhibit a general pattern that only depends on the values of the topological index $k$: for each $k$, there are $k$ single states, followed, in ascending order, by all other states which are nearly doubly degenerate. This energy level pattern reflects the fact that above the local minimum, states can be bound by both the local and global minima whereas below the local minimum states can only be bound by the global minimum. Since
the energy difference between the global and local minima increases  with $k$, the number of single states bound solely by the global minimum increases with $k$ as well (in fact is equal to $k$). States bound by both the global and local minima that lie below the maximum of the potential occur as nearly-degenerate doublets. All the intersections were found to be avoided, with the energy gaps exponentially decreasing with increasing field strengths.

{\it Supersymmetry.} We have identified two sets of conditions ({\it Case 1} and {\it Case 2}) under which SUSY QM can be applied to the 3D pendular eigenproblem. In particular, these conditions  imply a certain ratio of the interaction parameters $\eta$ and $\zeta$ and, thereby, a certain value of the topological index $k$. This made it possible to identify each {\it Case} with a particular topology: {\it Case 1} with $k=\mp m+1$, {\it Case 2} with $k=1$ for $m>0$.  In most cases SUSY QM enables us to obtain analytic expressions for the ground state wavefunctions. In Case $1_-$ for $m=1$, we were able to find the excited state wavefunction as well. The free rotor has been identified as a subcase of {\it Case $1_-$}, one which possesses analytic solvability for all states. 

Finally, we emphasize that the condition for analytic solvability as identified by SUSY coincides with the loci of the avoided eigensurface intersections, which only arise for the combined orienting and aligning interactions whose parameters $\eta$ and $\zeta$ span the eigensurfaces. So far, no SUSY analytic solution has been found for either of these interactions alone. 
The origin of the connection between eigensurface topology and exact solvability will be the subject of  a further investigation. 

\begin{acknowledgments}
Support by the DFG through grants  SCHM 1202/3-1 and FR 3319/3-1 is gratefully acknowledged.
\end{acknowledgments}

\bibliography{CombFields}

\clearpage

\begin{table}
	\centering
		\begin{tabular}{c|c|c|l|l|c|l|r}
		  \hline \hline
		   Case & $m$ & $k$ &  $V _1(\theta)+\beta^2\cos^2\theta+\tfrac{1}{4}$ & $\tilde{m}$ & $\tilde{k}$ & $V _2(\theta)+\beta^2\cos^2\theta+\tfrac{1}{4}$ & $\epsilon$ \\  \hline
			$1_\pm$,$2_\pm$  & 0  &  $1$ &  $-\tfrac{1}{4}\csc^2\theta - 2\beta\cos\theta$  & $1$ & $0$   & $\tfrac{3}{4}\csc^2\theta$ & $-\beta^2=-100$\\ \hline 
		$1_+$ & $1$ & $0$ &  $\tfrac{3}{4}\csc^2\theta$  & 0   & $-1$    & $-\tfrac{1}{4}\csc^2\theta  +2\beta \cos\theta$ & $-\beta^2=-100$\\ 
		$1_-$ & $1$   & $2$ &   $\tfrac{3}{4}\csc^2\theta -4\beta \cos\theta$  & $2$ & $1$      & $\tfrac{15}{4}\csc^2\theta  -2\beta \cos\theta$ & $2-\beta^2=-98$ \\  
		$2_+$  &	$1$  &  $1$ & $\tfrac{3}{4}\csc^2\theta - 2\beta\cos\theta$  &  -- & --  &  $\tfrac{7}{4}\csc^2\theta  -2\cot\theta\csc\theta$ & $-2\beta-\beta^2=-120$ \\  
		$2_-$  &	$1$  &  $1$ &  $\tfrac{3}{4}\csc^2\theta  - 2\beta\cos\theta$  & -- & --  & $\tfrac{7}{4}\csc^2\theta  +2\cot\theta\csc\theta$ & $2\beta-\beta^2=-80$ \\ \hline 
		$1_+$  &	$2$  &  $-1$ &  $\tfrac{15}{4}\csc^2\theta  +2\beta\cos\theta$  & $1$ & $-2$  & $\tfrac{3}{4}\csc^2\theta  +4\beta \cos\theta$ & $2-\beta^2=-98$ \\ 
			$1_-$ & $2$  &  $3$ &  $\tfrac{15}{4}\csc^2\theta -6\beta \cos\theta$  & $3$ & $2$  & $\tfrac{35}{4}\csc^2\theta  -4\beta \cos\theta$ & $6-\beta^2=-94$\\  
		$2_+$  &	$2$  &  $1$ & $\tfrac{15}{4}\csc^2\theta - 2\beta\cos\theta$  & -- & --   &  $\tfrac{19}{4}\csc^2\theta  -4\cot\theta\csc\theta$ & $-4\beta-\beta^2=-140$ \\  
			$2_-$ & $2$  &  $1$ &  $\tfrac{15}{4}\csc^2\theta  - 2\beta\cos\theta$  & -- & --  & $\tfrac{19}{4}\csc^2\theta  +4\cot\theta\csc\theta$ & $4\beta-\beta^2=-60$\\ \hline \hline
		\end{tabular}
	\caption{Superpartner potentials $V _1(\theta)$ and $V _2(\theta)$ and eigenenergies $\epsilon$ for Cases $1_{\pm}$ and $2_{\pm}$ as a function of the projection quantum number $0\le m \le 2$ and $\tilde{m}$. Also shown are the values of the topological index $k$ and $\tilde{k}$ for the two superpartner potentials. The numerical values of the eigenenergies correspond to $\beta=10$.}
	\label{tab:partners}
\end{table}

\clearpage

\end{document}